\documentclass[a4paper]{PoS}

\title{Theory overview of electroweak physics at hadron colliders}

\ShortTitle{Theory overview of electroweak physics at hadron colliders}

\author{\speaker{John Campbell}\\
        Fermilab, Batavia, IL 60510, USA\thanks{FERMILAB-CONF-16-336-T}\\
        E-mail: \email{johnmc@fnal.gov}}

\abstract{This contribution summarizes some of the important theoretical progress that has been made
in the arena of electroweak physics at hadron colliders.  The focus is on developments that have
sharpened theoretical predictions for final states produced through electroweak processes.  Special
attention is paid to new results that have been presented in the last year, since LHCP2015, as well
as on key issues for future measurements at the LHC.}

\FullConference{Fourth Annual Large Hadron Collider Physics\\
		13-18 June 2016\\
		Lund, Sweden}

\begin{document}

\section{Overview}

In Run 2 of the LHC a large portion of the experimental program
will be devoted to performing a detailed examination of the newly-discovered
Higgs boson and to probing the boundaries of the Standard Model (SM).
In this vein, precision studies of the mass, couplings
and decay modes of the Higgs boson, together with continued searches
for supersymmetry, other signs of New Physics and dark matter will be
paramount.  The vast amount of top quarks collected in the run will
also allow its production and decay characteristics to be scrutinized
like never before. In order to carry out this rich physics program it
will be essential to understand in great detail another class of
processes that can be loosely termed {\em electroweak}: the production
of vector bosons ($\gamma$, $W$- or $Z$-bosons) either singly, in pairs or at even
higher multiplicities, possibly in association with additional hadronic
activity (jets).  Such processes provide important, often irreducible,
backgrounds to the aforementioned measurements and searches.  These must be
understood and checked in great detail in order to best leverage data
from the LHC Run 2.  Beyond such pragmatic considerations, these processes
have their own intrinsic interest that spans from precision measurement of
SM parameters to direct tests of the nature of the electroweak sector.  Pinning-down
the parameters of the SM through measurements of the $W$-mass
and the weak mixing angle has been discussed elsewhere at this
conference\footnote{See the contribution of M. Schott, {\em ``Precision electroweak
physics at hadron colliders''}.}.  Related topics, such
as making improved determinations of parton distribution functions through
precise measurements of electroweak processes, have been touched on by many
speakers.  A different direction is represented by attempts to explore whether
the electroweak sector, as it is currently understood, is a complete
description of the theory.  This encompasses both searches for anomalous gauge
boson couplings and tests of the unitarity-cancellation mechanism in the
SM, which can be probed in both inclusive vector boson production
and in vector boson scattering~\cite{Lee:1977eg}.  Recent advances in
understanding the relevant theoretical challenges in addressing these
questions will be detailed in this contribution.
 
\begin{figure}
\begin{center}
\includegraphics[width=0.8\textwidth]{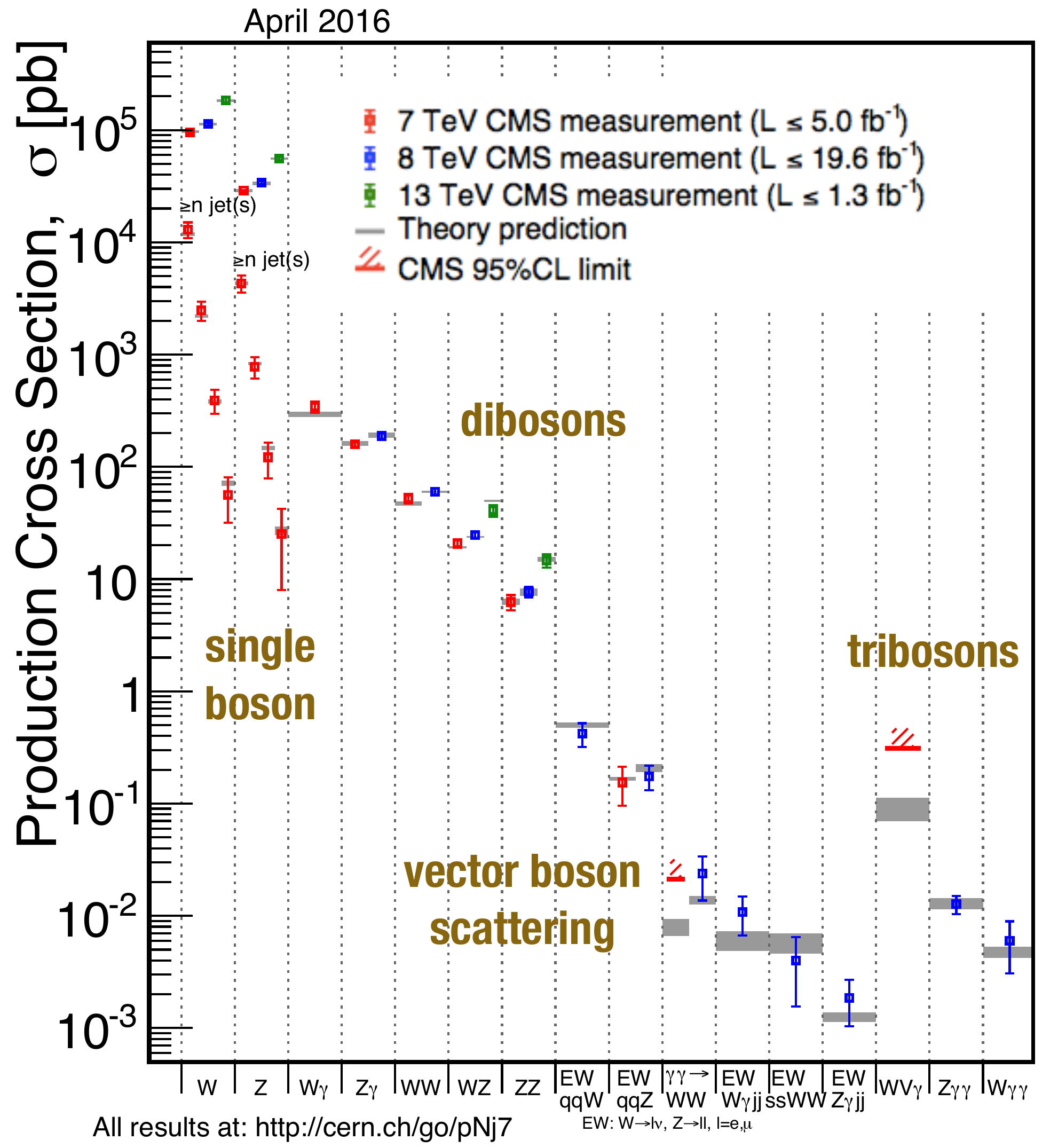}
\end{center}
\caption{Overview of electroweak cross-sections measured by the CMS experiment,
compared with theoretical predictions available in Spring 2016.  Minimally-adapted
from the CMS TWiki~\cite{CMSxsecs}.}
\label{Fig:CMSEWoverview}
\end{figure}

The plethora of measurements of electroweak cross-sections at the LHC, both 
in Run 1 and in the early results at $13$~TeV, have revealed a picture that
is very consistent with theoretical predictions of the SM
(Figure~\ref{Fig:CMSEWoverview}).  This impressive agreement, the result
of a great deal of concerted effort on both the experimental and theoretical
sides, raises an interesting conundrum.  The more this agreement persists,
the more precision is required on both sides in order to exclude subtle deviations
smaller than current uncertainties.  It is therefore important
to understand the current level of theoretical understanding of these processes,
their inherent limitations and the prospects for further improvement.  This
typically involves not only higher-order calculations in QCD, but also the
calculation of electroweak corrections.  The combination of these two effects,
together with their inclusion in fully-fledged event generators, will be
a recurrent theme here.  In addition, as the theoretical description becomes
more sophisticated, it is important to understand which remaining
approximations can be lifted and which of the outstanding uncertainties are
most important to control.

\section{Drell-Yan processes}

Due to their relatively simple nature, these processes have often been at the forefront of
theoretical sophistication.  Most recently, they have provided the testing-ground for
a variety of methods for combining the effects of
next-to-next-to-leading (NNLO) QCD corrections with a parton shower.
The resulting event generators, SHERPA+BlackHat based on the
UN${}^2$LOPS method~\cite{Hoeche:2014aia}, POWHEG using the MiNLO procedure with
DYNNLO~\cite{Karlberg:2014qua} and most recently Geneva(+Pythia) including also NNLL resummation
of zero-jettiness~\cite{Alioli:2015toa}\footnote{See the contribution of F. Tackmann,
{\em ``Drell-Yan production at NNLO+NNLL+PS in Geneva''}.},
will be indispensable tools
in the years to come.  Since these techniques are relatively new, it is particularly
important that a number of methods have been realized, with slightly different approaches
whose merits can be judged directly against LHC data in the future.

\begin{figure}
\begin{center}
\includegraphics[width=0.8\textwidth]{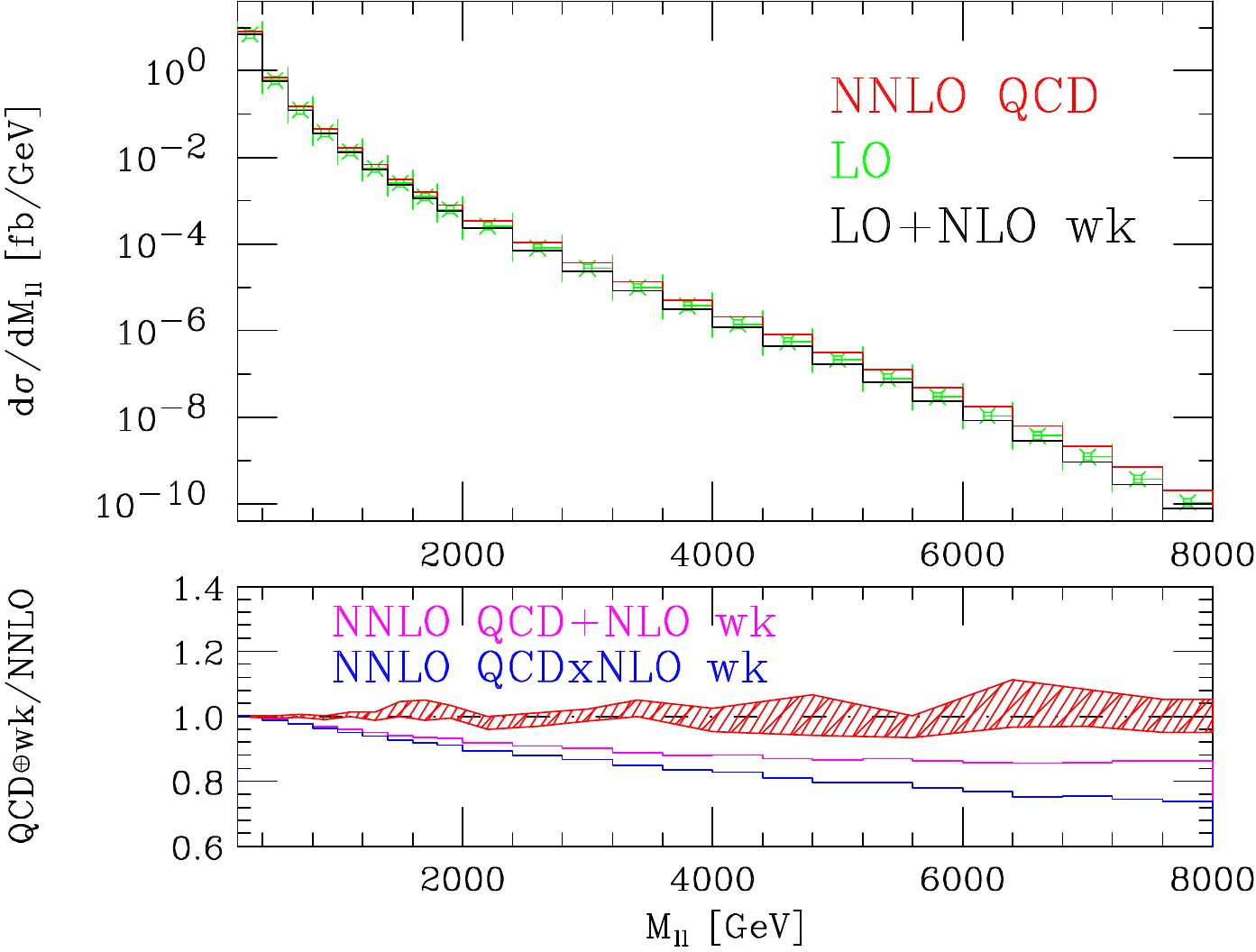}
\end{center}
\caption{The dilepton invariant mass distribution in $13$~TeV collisions predicted at LO (green),
NNLO QCD~\cite{Boughezal:2016wmq} (red, including scale variation) and with NLO weak
effects~\cite{Campbell:2016dks} (black).  The lower panel shows the difference between
the multiplicative (blue) and additive (magenta) combination procedures of NNLO QCD
and NLO weak effects (c.f. Eq.~(\protect\ref{eq:EWcomb})).}
\label{Fig:MCFMweak}
\end{figure}
At this level of precision it is mandatory to also include the effect of next-to-leading
order (NLO) electroweak corrections, that are naively of a similar size (since
$\alpha_s^2 \sim \alpha$).  Besides this simple argument, there are two distinct
kinematic regions that demand a treatment of electroweak effects.  The first is in
the region of resonant $W$ or $Z$ production where, in particular, corrections due to
the emission of real photons have a significant effect on the line-shape.
The second case where electroweak effects become important is in the region of
high invariant mass of the vector boson decay products, $\hat s$.
The calculation included in the Monte Carlo code FEWZ~\cite{Li:2012wna} accounts for 
corrections through both NNLO QCD and NLO EW and can be used to study the impact
of electroweak effects in both regions.  In the high-mass region the
one-loop electroweak corrections display an enhancement due to Sudakov logarithms whose
leading term corresponds to a factor $\log^2(\hat s/M_V^2)$, where $M_V$ is the vector boson
mass~\cite{Ciafaloni:1998xg}.  Defining the
individual QCD and weak higher-order corrections relative to the LO result by,
\begin{equation}
r^{QCD}(M) = \frac{d\sigma^{(NNLO~QCD)}/dM}{d\sigma^{LO}/dM} \,, \qquad
r^{wk}(M) = \frac{d\sigma^{(NLO~wk)}/dM}{d\sigma^{LO}/dM}  \,,
\end{equation}
there are two clear strategies for forming a combined correction.  These are,
\begin{equation}
r^{QCD\times wk} = r^{QCD} \times r^{wk} \qquad \mbox{and}, \quad
r^{QCD+wk} = r^{QCD} + r^{wk} - 1 \,.
\label{eq:EWcomb}
\end{equation}
The size of the individual corrections in the high-mass region means that the choice
of either of these procedures for combining NNLO QCD and NLO weak effects
gives rise to a significant ambiguity.   The difference between the two combinations
is outside typical NNLO QCD scale uncertainties
and reaches the $10\%$ level for very high dilepton invariant masses (Figure~\ref{Fig:MCFMweak}).
While such an uncertainty may be adequate given the current data set, it will be untenable
in the near future.

A way forward is of course to abandon any approximate combination, instead calculating exactly
the class of diagrams that provides a simultaneous enhancement by both couplings.  The first such
correction, of order $\alpha\alpha_s$, would correspond to a systematic improvement over combining
NLO calculations at each order.  Such a calculation entails the computation of two-loop diagrams
that contain loop propagators corresponding to both strong and weak particles.  Although the corresponding
master integrals are known~\cite{Bonciani:2016ypc}, a full calculation in this approach is not yet
possible.  However, results for the mixed QCD-EW corrections have been obtained very recently by
using the pole approximation~\cite{Dittmaier:2015rxo}.  Compared to a naive combination of QCD and EW
corrections, this calculation shows small but significant differences, for instance in the transverse
mass distribution of the $W$-boson.  This distribution is pivotal for an extraction of the 
$W$-mass and any change in the shape of the theoretical prediction leads to a systematic shift in the
extracted value of $M_W$.  The study of Ref.~\cite{Dittmaier:2015rxo} suggests that the mixed QCD-EW
corrections could lead to a change in the measured value of $M_W$ by as much
as $10$~MeV.  Since this is comparable to the precision that may be achieved in future at the LHC, it is
imperative that this effect be included.

\begin{figure}
\begin{center}
\includegraphics[width=0.46\textwidth]{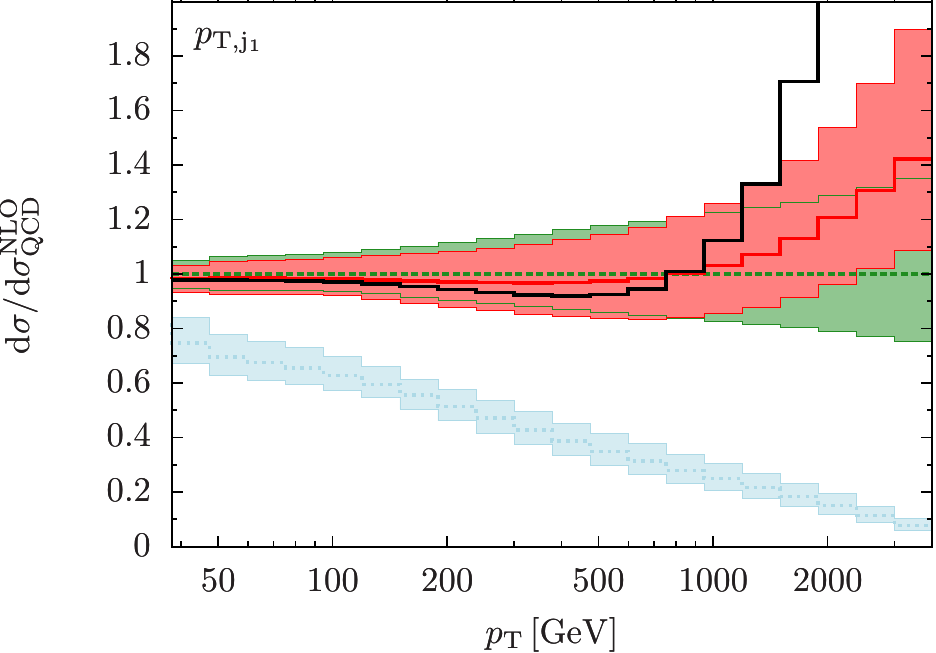}
\includegraphics[width=0.53\textwidth]{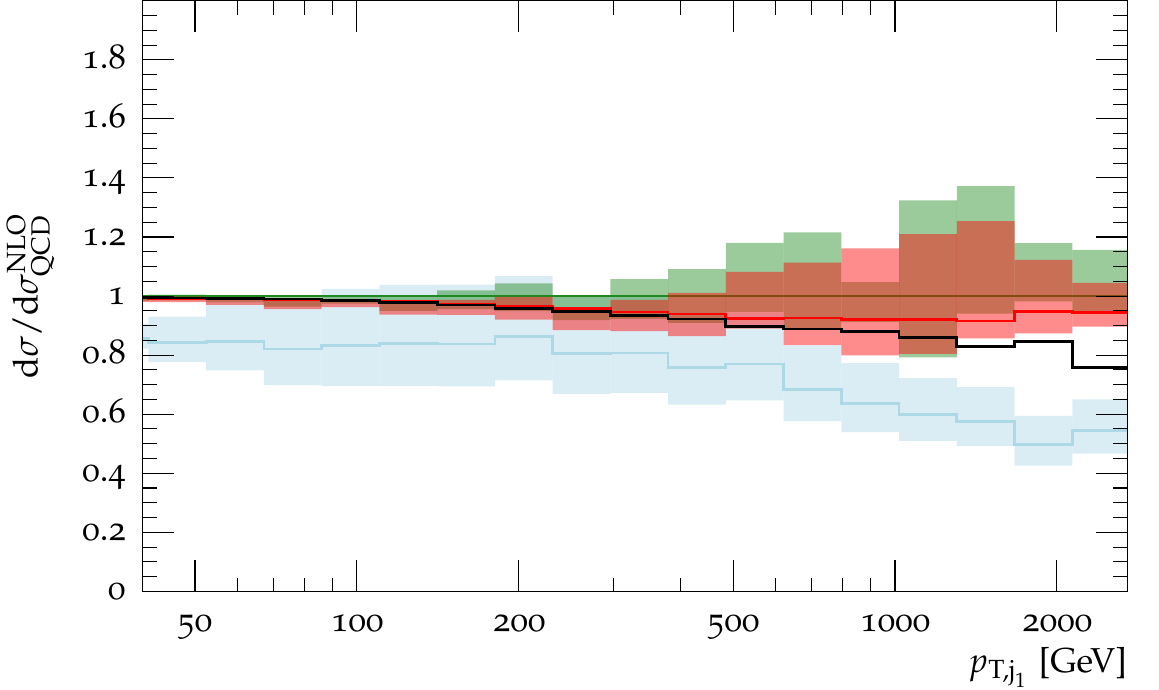}
\end{center}
\caption{The effect of combining NLO QCD and EW corrections on the lead jet transverse momentum distribution
in $W(\to \ell^- \bar\nu)$+jet production, in a fixed-order calculation (left) and after merging samples of
$V+1, 2$~jets using the MEPS procedure (right)~\cite{Kallweit:2015dum}.  The spread in the range of
predictions in the right-hand plot is much reduced compared to that on the left.}
\label{Fig:VjetEW}
\end{figure}

An alternative approach to addressing the issue of QCD and EW corrections simultaneously becoming large
can be provided by using a multijet merging procedure in a parton shower~\cite{Kallweit:2015dum}.\footnote{
See the contribution of M. Schoenherr, {\em ``NLO QCD+EW for V+jets''}.}
In this calculation, electroweak and QCD effects are 
computed for both $V+1$ and $V+2$~jet final states, then the results are combined using the usual
SHERPA merging of samples.   Compared to a strictly fixed-order approach, the ambiguity associated with
the combination of QCD and EW corrections is greatly reduced, to a much more palatable level
(Figure~\ref{Fig:VjetEW}).

\begin{figure}
\begin{center}
\includegraphics[width=0.7\textwidth]{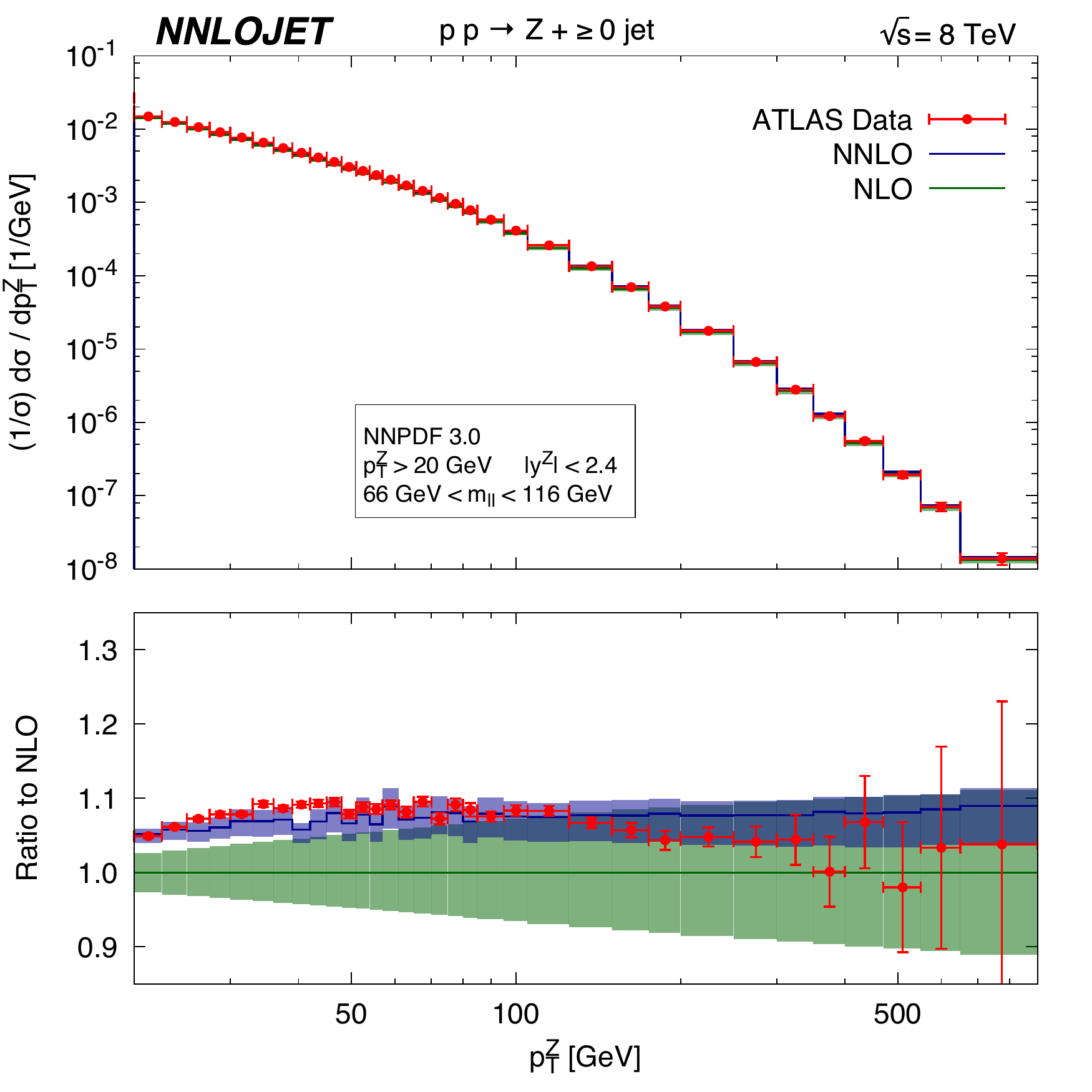}
\end{center}
\caption{Comparison of the $Z$-boson transverse momentum through NNLO QCD with data taken
at $8$~TeV by the ATLAS experiment~\cite{Ridder:2016nkl}.}
\label{Fig:GdRZjet}
\end{figure}

One of the significant theoretical breakthroughs in perturbative QCD in the last year has been the
ability to perform NNLO QCD calculations of V+jet processes.  An important cross-check of the calculations
is that they have been performed using both antenna subtraction~\cite{Ridder:2015dxa,Ridder:2016nkl} and
jettiness subtraction~\cite{Boughezal:2015dva,Boughezal:2015ded,Boughezal:2016yfp}.
The NNLO calculations enable a much-improved theoretical description of, for instance, the transverse
momentum distributions of the $Z$-boson and the associated jet.  
The importance of such distributions has been emphasized more than once at this meeting\footnote{See the
contributions of G. Salam, {\em ``Theory overview of QCD physics at hadron colliders''}
and R. Boughezal, {\em ``Developments in QCD high order calculations''}.},
which is due to the fact that experimental measurements
can be made with percent-level precision up to transverse momenta of about $200$~GeV.
At present some tension with ATLAS Run I data remains~\cite{Ridder:2016nkl}, even after
normalizing by the inclusive $Z$-boson cross-section in order to remove
ambiguities associated with the uncertainty in the total luminosity (Figure~\ref{Fig:GdRZjet}).
The combination of such exquisite data with the sophisticated NNLO predictions, with uncertainties
that are only a little larger, offers the opportunity for future parton distribution function constraints in this channel.

\section{Diboson production}

At the LHCP conference in 2015, theoretical predictions for diboson production were already very
advanced, with significant breakthroughs in recent years allowing the computation of many of these processes
in NNLO QCD.  In the intervening year yet more progress has been made, with NNLO QCD and NLO EW corrections
now computed for all processes.  In addition, various refinements -- removing previous approximations that had been used
-- have been implemented, together with first results for a class of important contributions that enters
in N${}^3$LO QCD.  The situation is summarized in Table~\ref{Table:DibosonSummary}. 
\begin{table}
\begin{center}
\begin{tabular}{|l|l|l|l|}
\hline
Process & LHCP 2015 status  & New developments & Comments \\
\hline
 $\gamma\gamma$ & NNLO QCD~\cite{Catani:2011qz} & NNLO(+) QCD~\cite{Campbell:2016yrh} 
 & independent calculation; \\
 & & & improved treatment of \\
 & & & $gg$ contribution \\
\hline
$V\gamma$ & NNLO QCD~\cite{Grazzini:2015nwa} &  & includes off-shell effects, \\
& NLO EW ($W\gamma$)~\cite{Denner:2014bna} & NLO EW ($Z\gamma$)~\cite{Denner:2015fca} & radiation from leptons \\
& & & in decay \\
\hline
$WW$ & NNLO QCD~\cite{Gehrmann:2014fva} & NNLO QCD~\cite{Grazzini:2016ctr} & single-resonant diagrams, \\
&NLO EW (approx.)  & beyond NNLO QCD~\cite{Caola:2015rqy} & same-flavor lepton $ZZ$\\
& & NLO EW~\cite{Biedermann:2016guo} & contributions; higher-order \\
& & & $gg$ loops, exact EW \\
\hline
$WZ$ & NLO QCD & NNLO QCD~\cite{Grazzini:2016swo} & off-shell effects, \\
&NLO EW (on-shell)  &  & single-resonant diagrams \\
\hline
$ZZ$ & NNLO QCD~\cite{Grazzini:2015hta} & beyond NNLO QCD~\cite{Caola:2015psa} & higher-order $gg$ loops, \\
&NLO EW (approx.)  & NLO EW~\cite{Biedermann:2016yvs} & exact EW \\
\hline 
\end{tabular}
\caption{Summary of the status of theoretical calculations of diboson production at the
time of LHCP 2015 and the developments since then.}
\label{Table:DibosonSummary}
\end{center}
\end{table}

\begin{figure}
\begin{center}
\includegraphics[width=0.9\textwidth]{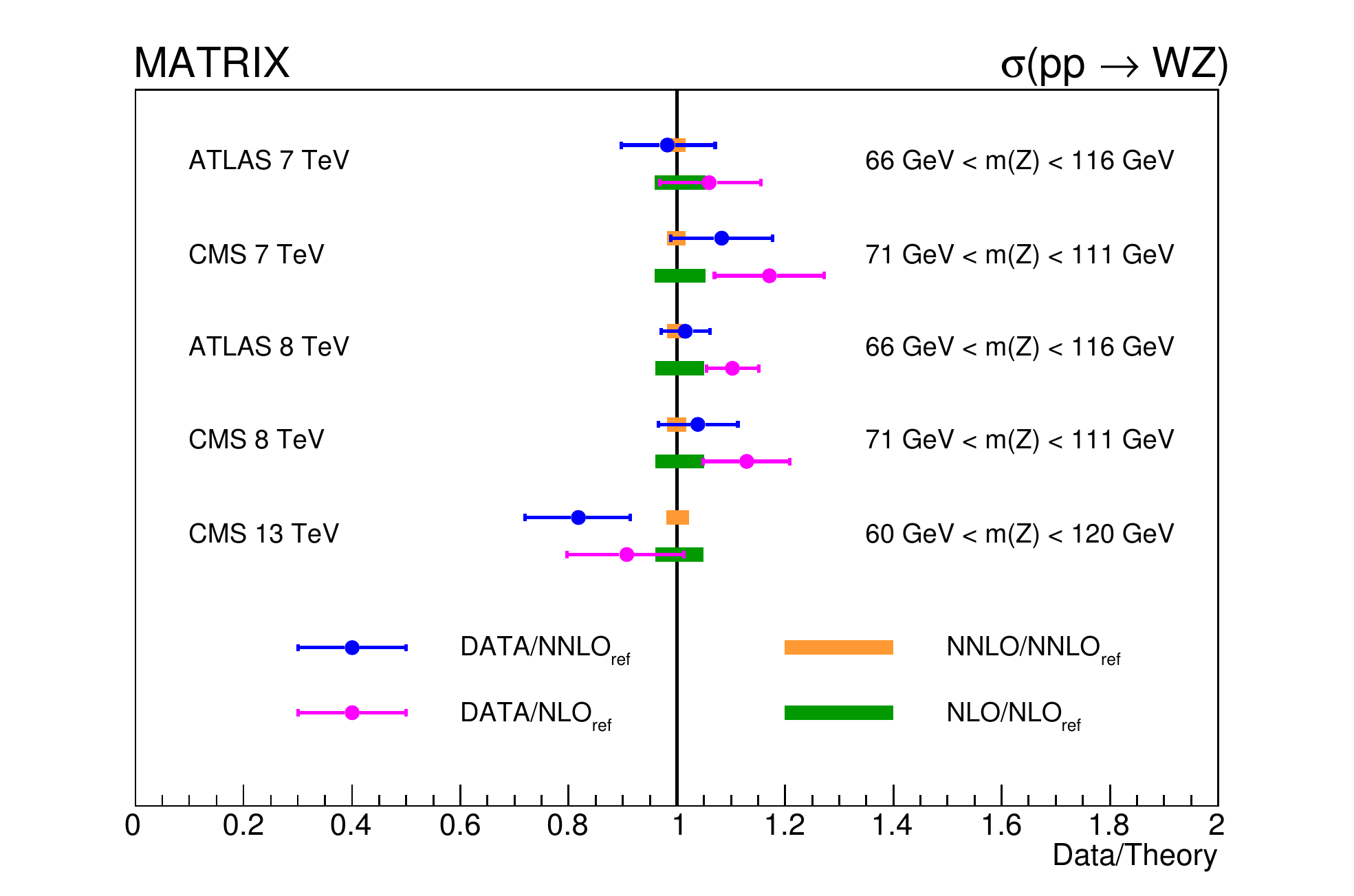}
\end{center}
\caption{A NNLO calculation of $WZ$ production, compared with experimental measurements performed by
ATLAS and CMS at various LHC operating energies~\cite{Grazzini:2016swo}.}
\label{Fig:WZnnlo}
\end{figure}

Many of the new diboson NNLO calculations in the last year have been performed using a general
framework called MATRIX~\cite{Grazzini:2016swo,Grazzini:2016ctr}\footnote{
See the contribution of S. Kallweit, {\em ``NNLO di-boson production''}.}.
In some cases the new calculations have extended the applicability
of previous results, for instance by including off-shell effects in the vector boson decays and
accounting for the contribution of single-resonant diagrams.  This can have important consequences, for instance
on the acceptance (the ratio of the fiducial to inclusive cross-section) predicted by theory.  
In the case of $WW$ production, the acceptance decreases by approximately $3$\% when moving from NLO
to NNLO, for typical LHC analyses at both $8$ and $13$~TeV~\cite{Grazzini:2016ctr}.
The missing diboson process, $WZ$ production, was also
completed this year~\cite{Grazzini:2016swo} and agrees well with LHC data taken at a variety of
energies (Figure~\ref{Fig:WZnnlo}).

\begin{figure}
\begin{center}
\begin{tabular}{lr}
\begin{minipage}{0.45\textwidth}
\includegraphics[width=\textwidth]{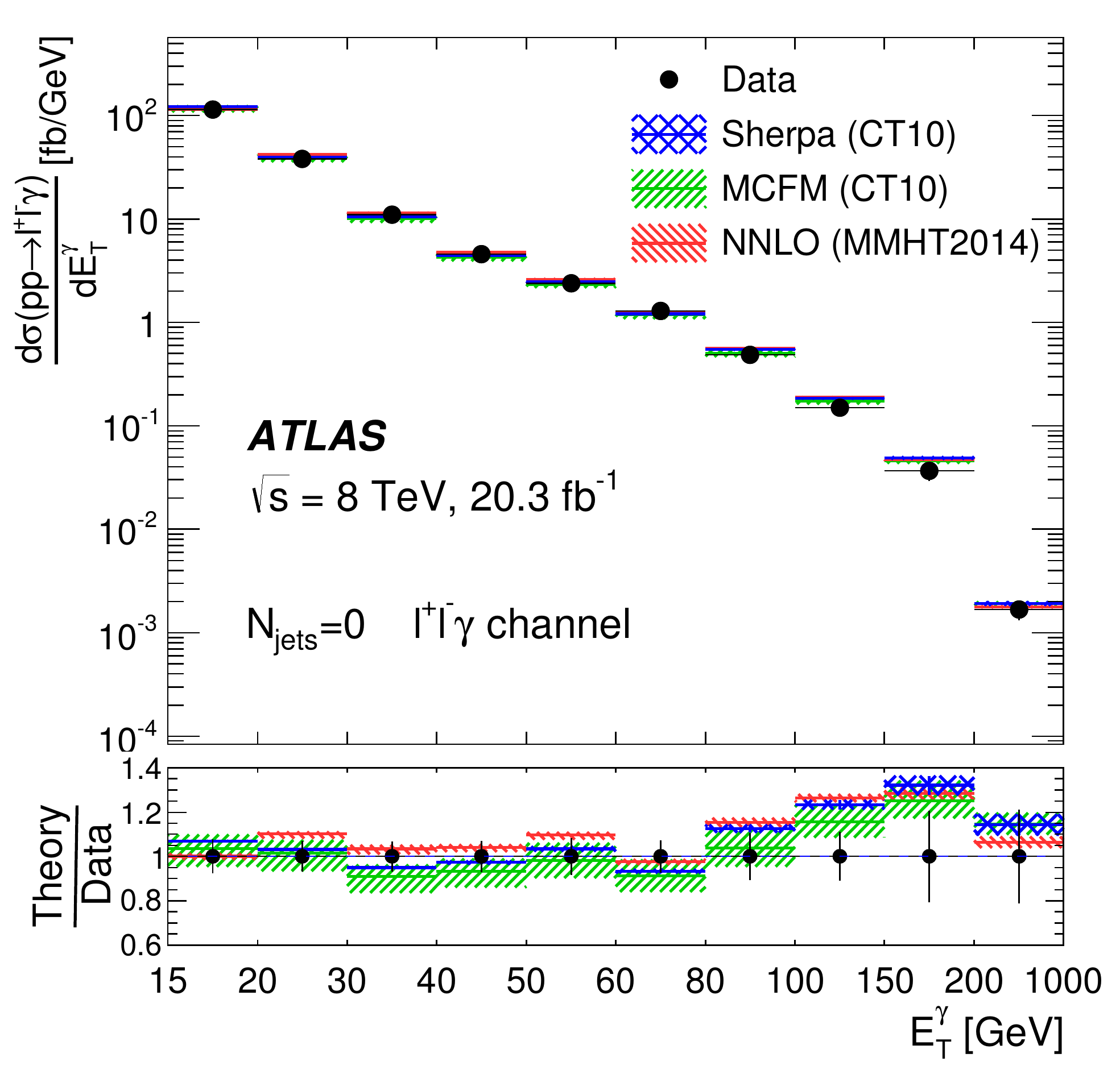}
\end{minipage} &
\begin{minipage}{0.55\textwidth}
\includegraphics[width=\textwidth]{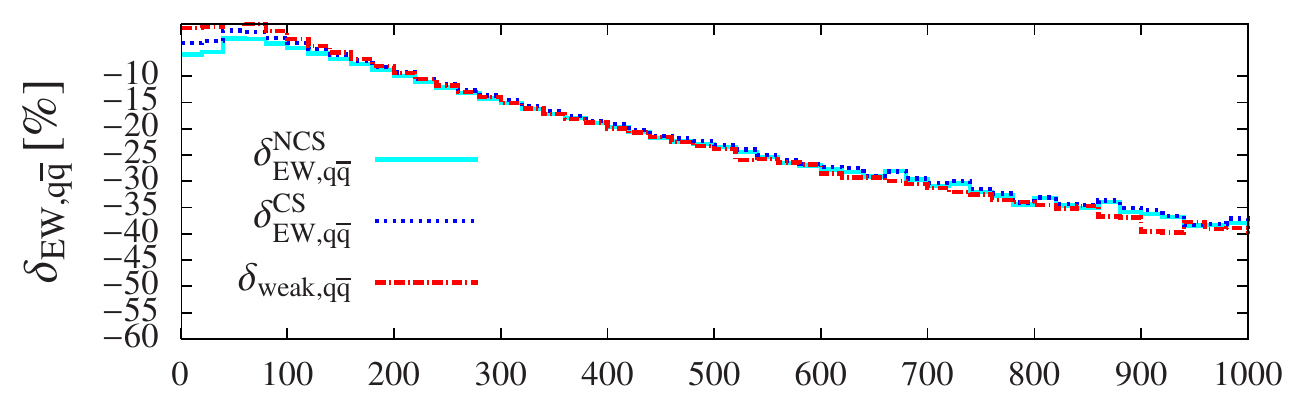} \\
\includegraphics[width=\textwidth]{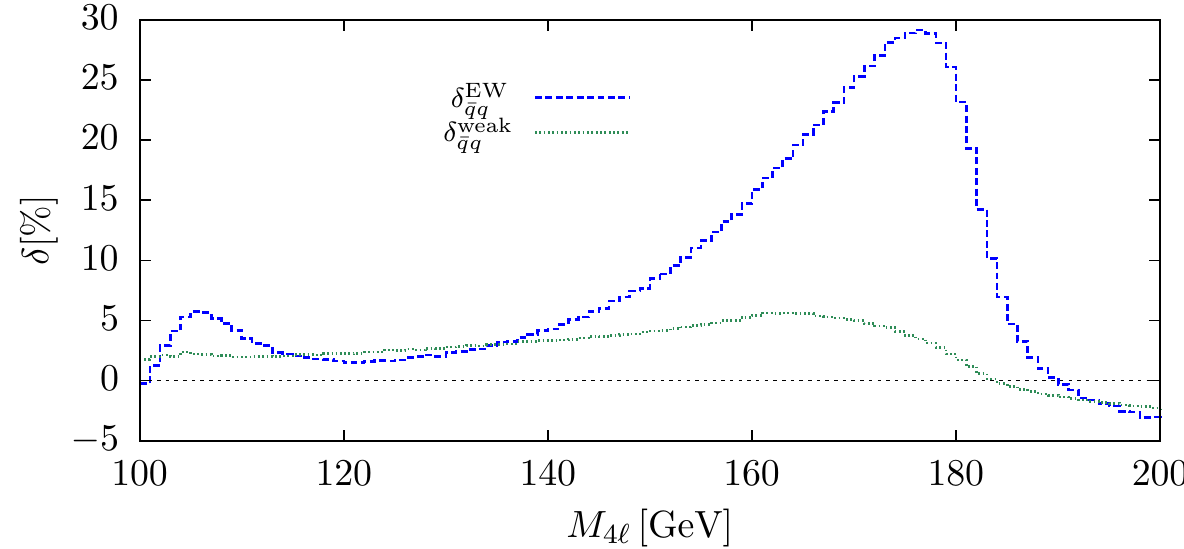}
\end{minipage}
\end{tabular}
\end{center}
\caption{Illustrations of the impact and importance of NLO electroweak corrections to
diboson processes: (left) an ATLAS analysis of $\ell^+\ell^-\gamma$ events~\cite{Aad:2016sau};
(upper right) EW effects in the same channel~\cite{Denner:2015fca};
(lower right) EW effects in $ZZ$ production~\cite{Biedermann:2016yvs}.}
\label{Fig:EWdibosons}
\end{figure}

Full NLO electroweak corrections have also recently been presented for the cases of $Z\gamma$, $WW$ and
$ZZ$ production.  In each case important phenomenological consequences of the results of these
calculations have been observed.  As a first example, consider ATLAS data taken in the $Z\gamma$ channel
during Run 1 of the LHC~\cite{Aad:2016sau} (Figure~\ref{Fig:EWdibosons}, left), which shows that there is ample data for
photon transverse momenta up to about $200$~GeV.  Comparing with the calculation of NLO EW effects up to
this momentum scale (Figure~\ref{Fig:EWdibosons}, upper right), it is clear that they are rather mild
(as large as $-10$\%) in Run 1~\cite{Denner:2015fca}.
However this will no longer be the case in Run 2, where cross-sections and data samples will be
much larger and higher transverse momenta will be probed.  An additional observation is that, in some
cases -- for instance after application of a jet veto, as in the example at hand -- the NLO EW corrections can become the leading
effect, larger than the effect of NLO QCD.  A final example underscores once again the importance of
a calculation that extends beyond the on-shell case, to full off-shell accuracy (Figure~\ref{Fig:EWdibosons},
lower right).  In this case such a calculation of the $ZZ$ cross-section shows that the behaviour of the
EW corrections is very different (opposite in sign) between the on-shell ($M_{4\ell} \sim 2M_Z$) and
Higgs search ($M_{4\ell} \sim M_H$) regions~\cite{Biedermann:2016yvs}.

\begin{figure}
\begin{center}
\includegraphics[width=0.5\textwidth]{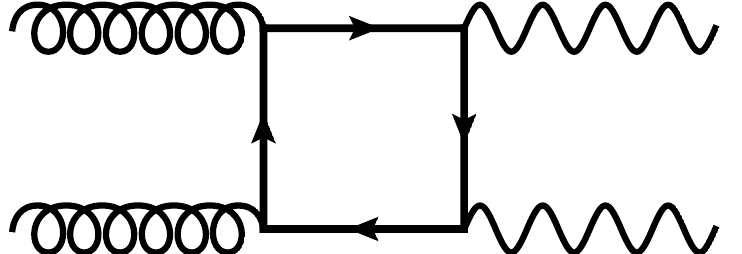}
\end{center}
\caption{A one-loop diagram representing the process $gg \to V_1 V_2$ that enters the NNLO QCD
calculation of $V_1 V_2$ production.}
\label{Fig:nfloop}
\end{figure}

One of the interesting aspects of the theoretical predictions for gauge boson pair production
at NNLO is the sensitivity to $gg$ initial states in the neutral channels, i.e.
$gg \to W^+W^-$,~$ZZ$ or~$\gamma\gamma$.  Such contributions enter through diagrams containing a closed
loop of quarks, as shown in Figure~\ref{Fig:nfloop}.  Since the SM contains no tree-level
coupling of the form $gg V_1 V_2$ it is clear that, despite representing one-loop contributions, these diagrams
are finite.  The resulting impact on the total cross-section is small, dwarfed by initial states
containing quarks, but is important at the level of the corrections that enter at NNLO.  In fact these
represent the most important NNLO effect in the case of $ZZ$ production~\cite{Grazzini:2015hta}.
Given this importance at NNLO it is worthwhile to consider whether this contribution, which is included
for the first time and therefore suffers all the foibles of a leading order calculation, could be computed
at the next order of perturbation theory.

\begin{figure}
\begin{center}
\includegraphics[width=0.54\textwidth]{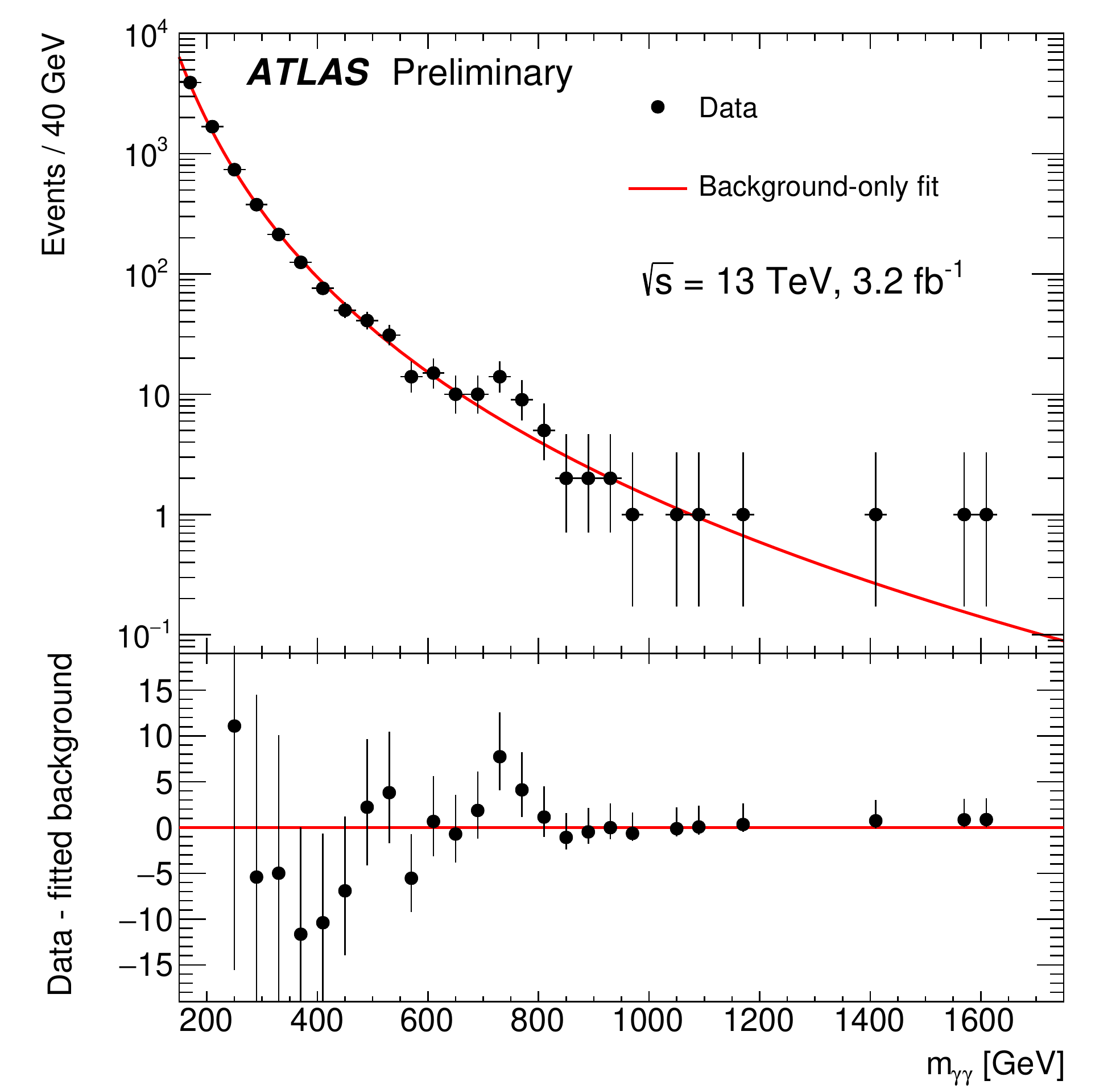}
\includegraphics[width=0.44\textwidth]{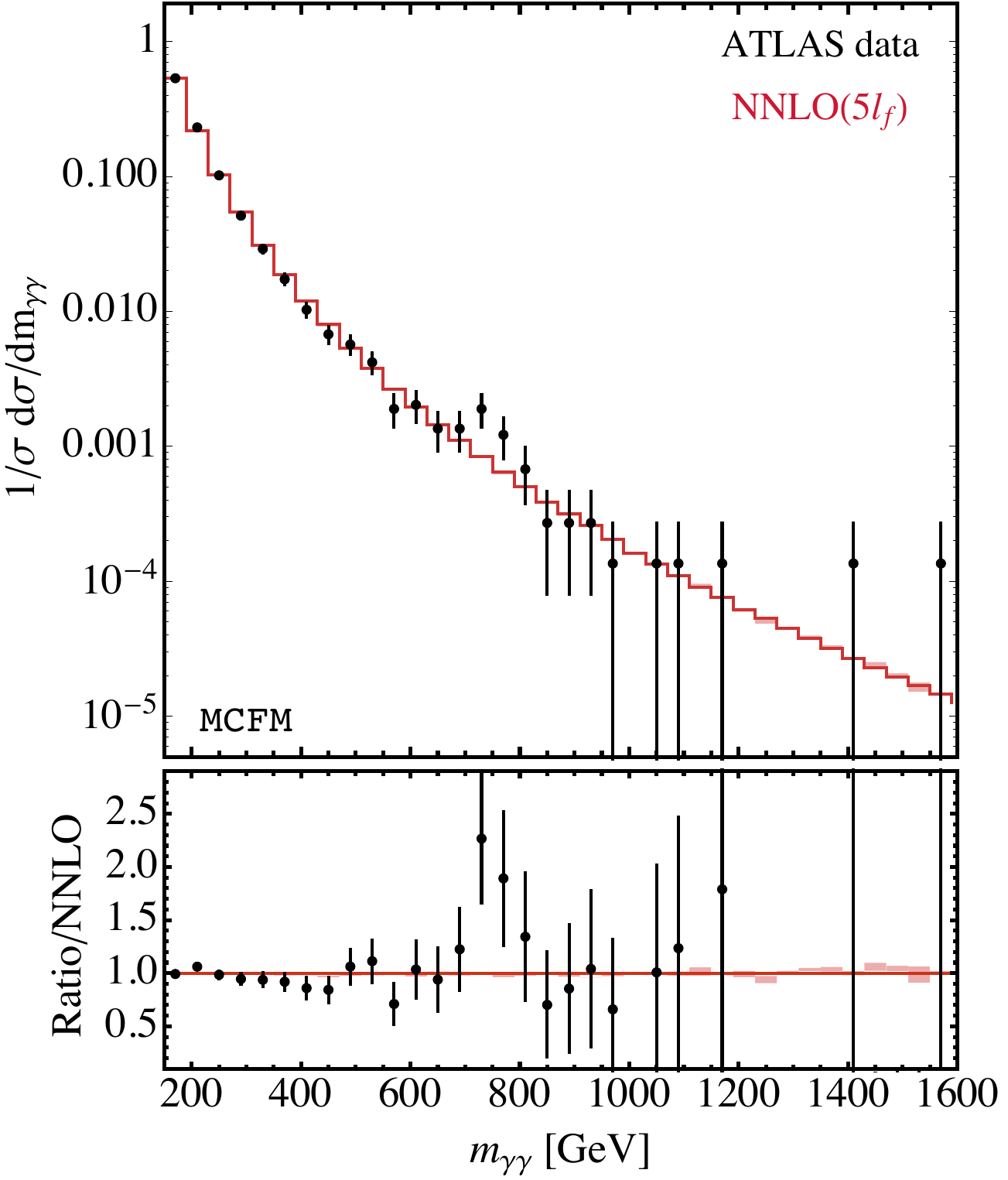}
\end{center}
\caption{A comparison of diphoton data from the ATLAS experiment and accompanying phenomenological
fit~\cite{ATLAS-CONF-2015-081} (left) with a NNLO calculation~\cite{Campbell:2016yrh} (right).}
\label{Fig:diphotonMgg}
\end{figure}

Corrections at the next order 
(NLO) for the $gg$-induced components (i.e. entering the full calculation at N${}^3$LO), were first considered
for the case of light quark loops in diphoton production~\cite{Bern:2001df,Bern:2002jx}.  A recent independent
calculation of the diphoton process at NNLO has taken into account the leading effect of heavy quark loops
and also included the partial N${}^3$LO effects for loops of light quarks~\cite{Campbell:2016yrh}.  The partial
N${}^3$LO effects can have as large as a $5$-$10$\% effect on the cross section at small diphoton invariant masses
and they slightly improve the description of 7~TeV data presented by the CMS experiment~\cite{Chatrchyan:2014fsa}.
The top quark loops have a smaller effect and sculpt the shape of the $m_{\gamma\gamma}$ distribution in the
$300 - 500$~GeV region.  However the effects are far too small to significantly affect the description of the
$m_{\gamma\gamma}$ distribution in the region of $750$~GeV, where both ATLAS and CMS had indicated a slight
excess at the beginning of Run 2 of the LHC~\cite{ATLAS-CONF-2015-081,CMS-PAS-EXO-15-004}.  In fact the NNLO
calculation provides an excellent {\em ab initio} calculation of this distribution that agrees very well with
the fitting form that had been used by ATLAS and CMS, albeit without any attempt to account for fake rates or
efficiencies.  A comparison of the ATLAS data and fitting form with the NNLO prediction is shown in
Figure~\ref{Fig:diphotonMgg}.

The effect of NLO corrections to the $gg$ contribution to $W^+ W^-$ and $ZZ$ final states is considerably
harder to compute due to the non-zero $W$ and $Z$ boson masses.  Nevertheless, such calculations have been
completed in the last year, including the effect of the vector boson decays~\cite{Caola:2015psa,Caola:2015rqy}.
In both cases the calculations have, in the first instance, been limited to loops that do not
contain the top quark (thus neglecting the entire third generation for $W^+ W^-$ production).  The effect
of the corrections is illustrated in Table~\ref{Table:ggVVnlo}.  The results for $ZZ$ production are shown
in the case when no cuts are applied on the $Z$ decay products.  In this case the $gg$ contribution increases by
almost a factor of two at NLO, but the effect on the overall rate is much smaller, about a $5\%$ enhancement.  For
$W^+ W^-$ production the cross-sections are shown after the application of fiducial cuts (taken from Ref.~\cite{Aad:2016wpd}),
which include a veto on additional jet activity.  This greatly reduces the effect of the NLO corrections in this case,
resulting in a negligible change in the best theoretical prediction for the cross-section.  Note that, even in the case
of $ZZ$ production where the impact of the $gg$ loops is biggest, the best measurements from Run 1 suffer from approximately
$10\%$ uncertainties~\cite{ATLAS:2013gma,CMS:2014xja} and thus are not yet sensitive to such corrections.
\begin{table}
\begin{center}
\begin{tabular}{|l||l|l||l|l|}
\hline
Final-state (cuts) & $\sigma_{gg}^{LO}$~~~~~~~~ & $\sigma_{gg}^{NLO}$~~~~~~~ &
 $\sigma^{NNLO}$~~~~~~~ & $\sigma^{NNLO} + \Delta\sigma_{gg}^{NLO}$ \\
\hline
$ZZ$ (no cuts) & 0.53\,{\mbox pb} & 0.95\,{\mbox pb} & 8.28\,{\mbox pb} & 8.70\,{\mbox pb} \\
$W^+ W^-$ (fiducial cuts)~\cite{Aad:2016wpd} & 9.8\,{\mbox fb} & 11.8\,{\mbox fb} & 355\,{\mbox fb} & 357\,{\mbox fb} \\
\hline
\hline 
\end{tabular}
\caption{The effect of NLO corrections to $gg$ contributions in vector boson pair production.
The final column shows the prediction that includes the full NNLO result and the NLO corrections
to the $gg$ channel, with $\Delta\sigma_{gg}^{NLO} = \sigma_{gg}^{NLO}-\sigma_{gg}^{LO}$.}
\label{Table:ggVVnlo}
\end{center}
\end{table}

\section{Off-shell Higgs boson studies}

One of the prime modes for making precision measurements of properties of the Higgs
boson is the process $gg \to H \to ZZ^* \to 4\ell$.  This combination of production and decay
modes leads to a substantial cross-section that is under good experimental control.
On the theoretical side there has been renewed interest in a precision calculation
of this process in the region where the Higgs boson is significantly off-shell.  This
was first motivated by the fact that a significant fraction, up to 15\%, of such events
occur in the region $m_{4\ell} > 2m_Z$~\cite{Kauer:2012hd}.  Further interest was
provided by the realization that, in the SM, the ratio of the off-shell
and on-shell rates is approximately proportional to the width of the Higgs boson~\cite{Caola:2013yja}.
In order to make use of this observation it is clearly imperative to have precise control
of the theoretical prediction.  In the off-shell region this means not
only the calculation of the diagrams involving the Higgs boson, but also of the $gg \to ZZ \to 4\ell$
quark loops such as the one in Figure~\ref{Fig:nfloop}.  These diagrams enter at the
same order in perturbation theory and make a significant contribution.  They interfere
destructively and, in the case of the top quark loops, this is essential in order to
render the high-energy behavior sensible~\cite{Glover:1988rg}.

Unfortunately, the calculation of the off-shell rate through $gg$ initial states is
complicated by the fact that it occurs at leading order through a one-loop process.  
As discussed already, the leading order nature of this contribution means that it suffers from substantial
scale dependence.  This means that, for instance, constraints on the Higgs boson width
or generic off-shell couplings are weakened by significant theoretical uncertainty~\cite{Campbell:2013una}.
As noted above, the calculation of higher-order corrections, that
should include loops of top quarks, is beyond the reach of current technology.  In the
absence of such predictions, a reasonable assumption is that the higher-order corrections
to loop diagrams such as the one in Figure~\ref{Fig:nfloop} should behave like the corrections
to the Higgs diagrams alone (that
are, of course, known).  This is the basis for a pragmatic estimate of limits on the
Higgs boson width as a function of the unknown NLO $K$-factor~\cite{Aad:2015xua}.

\begin{figure}
\begin{center}
\includegraphics[width=0.49\textwidth]{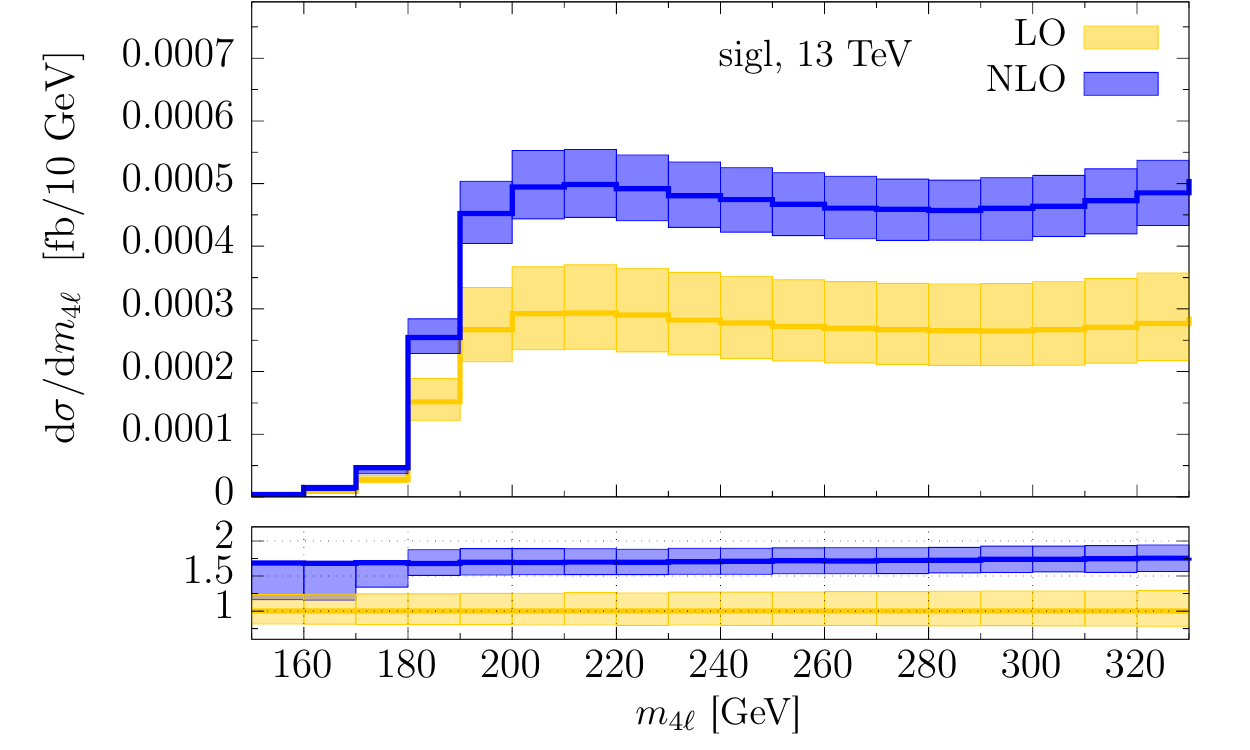}
\includegraphics[width=0.49\textwidth]{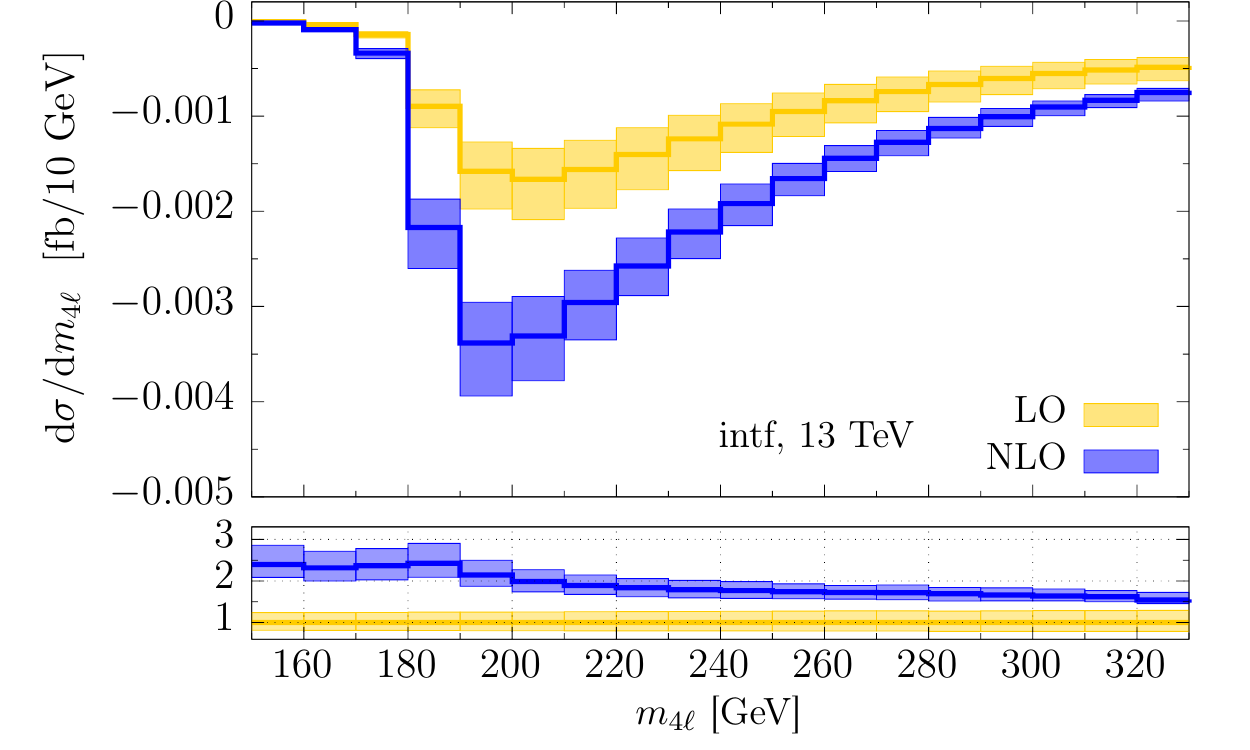}
\end{center}
\caption{ The four-lepton invariant mass distribution for $gg \to ZZ$ contributions
from the Higgs diagrams alone (left) and interference with other one-loop contributions (right).
Predictions taken from Ref.~\cite{Caola:2016trd}.}
\label{Fig:ggKfac}
\end{figure}

Although an exact calculation of higher-order corrections involving top quark loops
is not yet available, the last year has seen significant progress towards this goal.
From approximations based on soft gluon resummation~\cite{Li:2015jva} to those based on an
expansion in inverse powers of the top quark mass~\cite{Melnikov:2015laa,Caola:2016trd}
(possibly including further refinements~\cite{Campbell:2016ivq}), a consensus is emerging. 
All of these calculations suggest that, in the region of interest, the $K$-factor for these
contribution is indeed very similar to the one for the Higgs diagrams alone.
This is shown, for example, in Figure~\ref{Fig:ggKfac}, taken from Ref.~\cite{Caola:2016trd}.
Although the $K$-factors (shown in the lower panels) are significantly different close to the
$2M_Z$ threshold -- a fact that appears to be due to the contribution of loops of massless
quarks -- in the highest energy
bins they are very similar.  This leads to a ratio of off-shell to on-shell production at NLO
that is very close to the ratio obtained at LO, except of course with much smaller scale
uncertainty~\cite{Campbell:2016ivq}.

\section{Beyond inclusive diboson production}

Although the inclusive production of dibosons provides a detailed test of perturbative QCD
and has enabled stringent limits on anomalous triple gauge boson couplings to be placed,
as the LHC collects more
data it will become increasingly sensitive to new mechanisms of vector boson production.
These relatively-unexplored channels represent either less inclusive final states or
the production of more than two vector bosons.

\begin{figure}
\begin{center}
\begin{tabular}{ccc}
\includegraphics[width=0.27\textwidth]{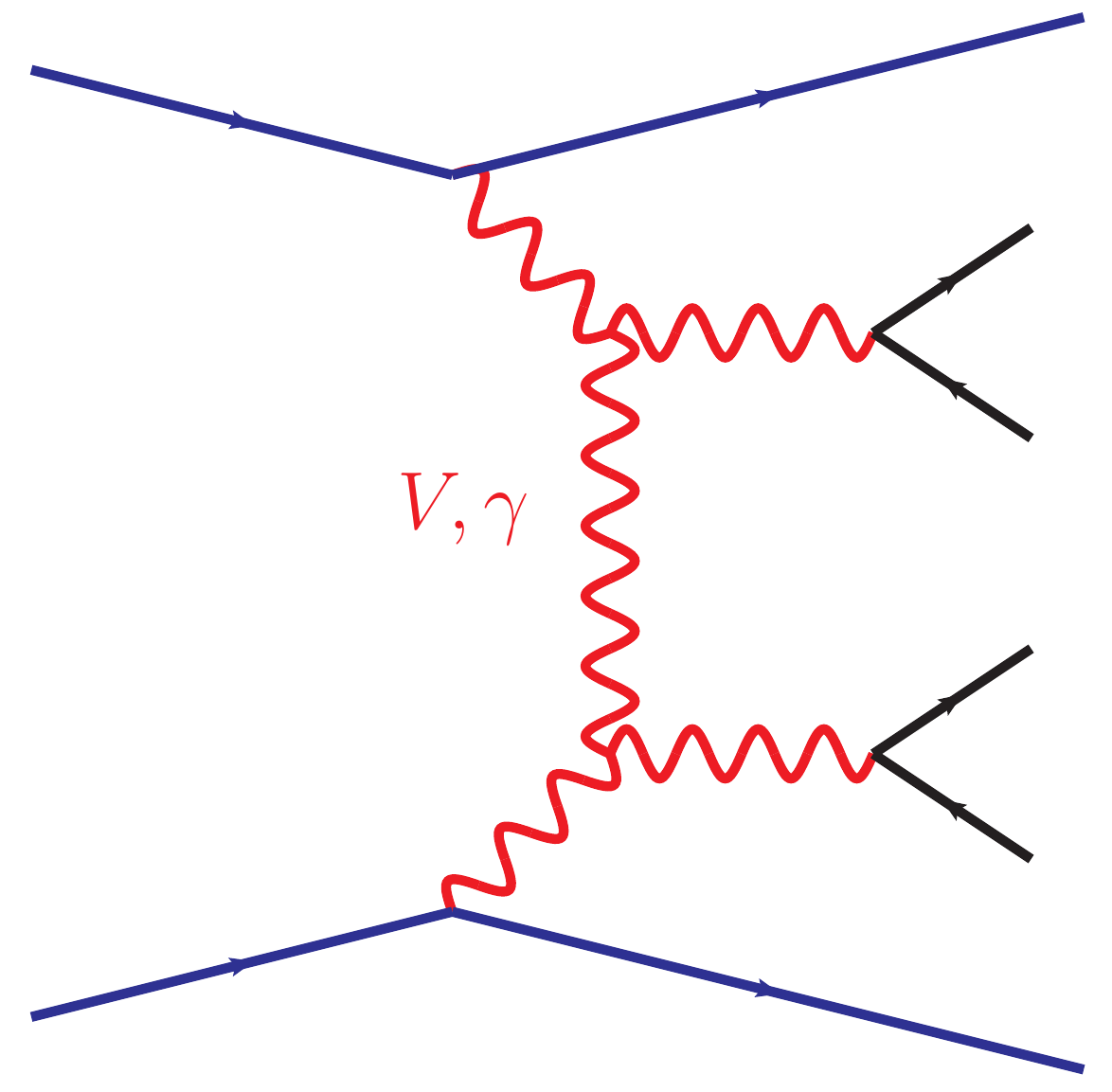} \hspace*{0.5cm} &
\includegraphics[width=0.27\textwidth]{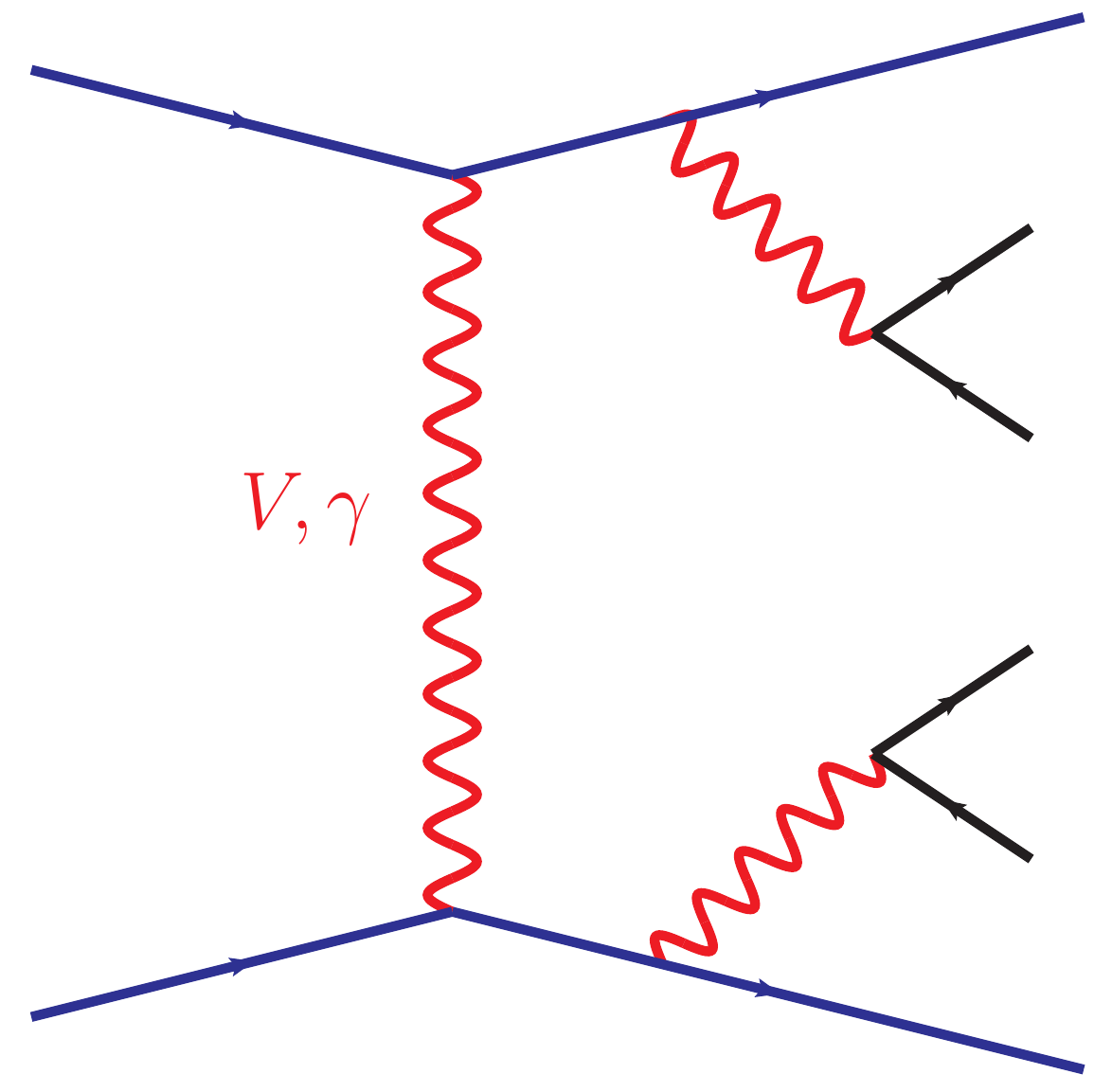} \hspace*{0.5cm} &
\includegraphics[width=0.27\textwidth]{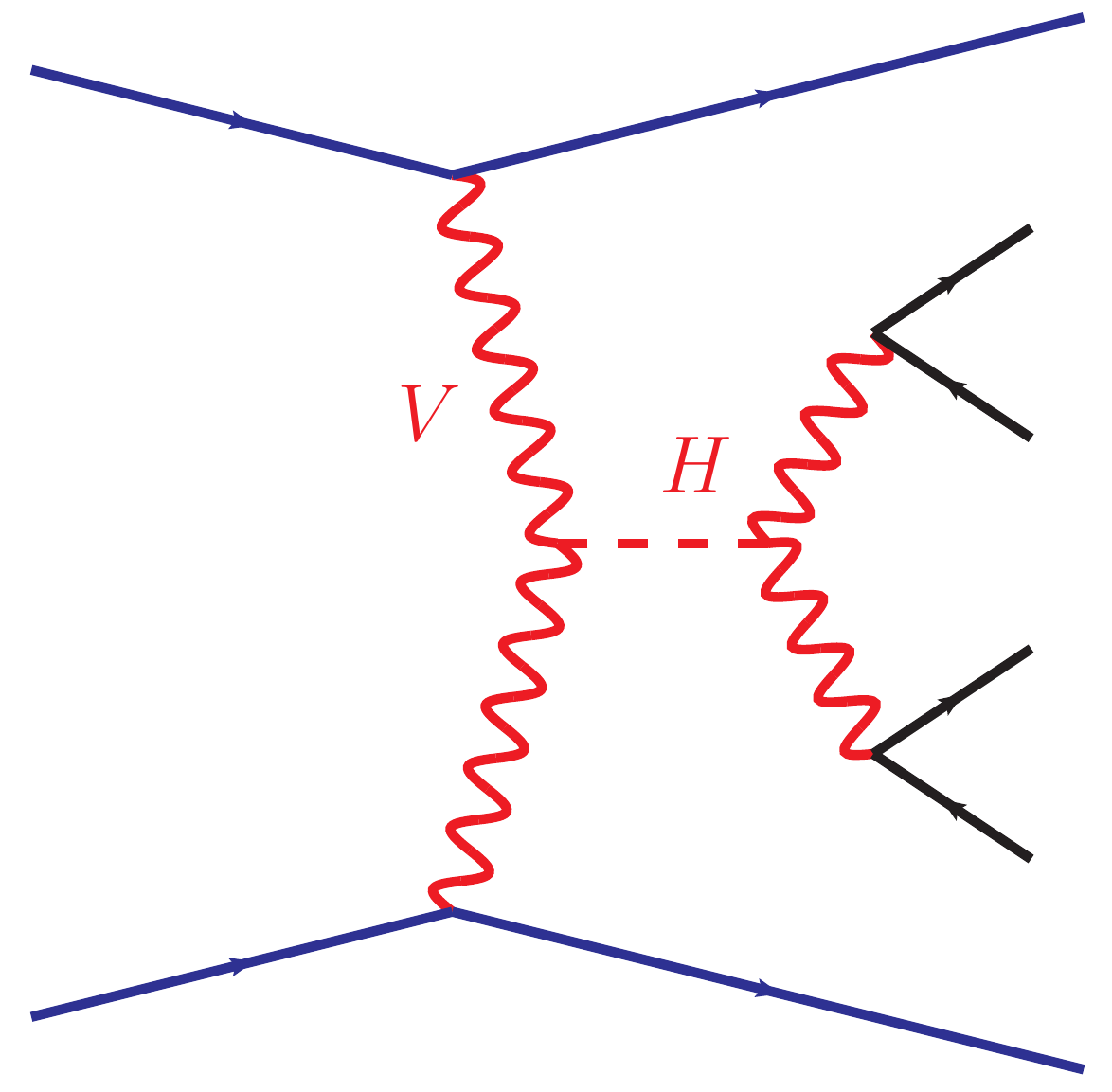} \\ 
(a) & (b) & (c) \\ & & \\
\includegraphics[width=0.27\textwidth]{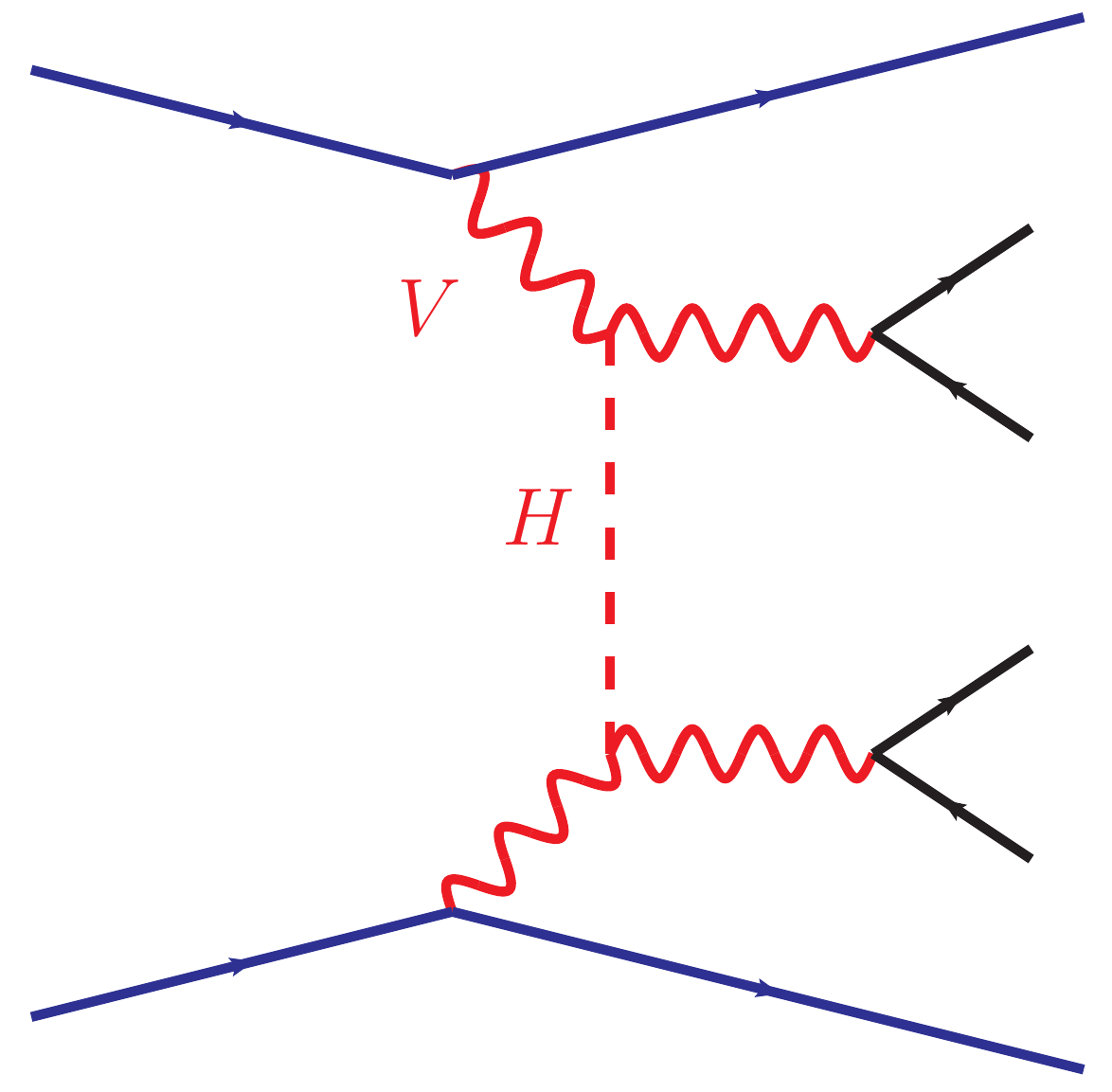} &
\includegraphics[width=0.27\textwidth]{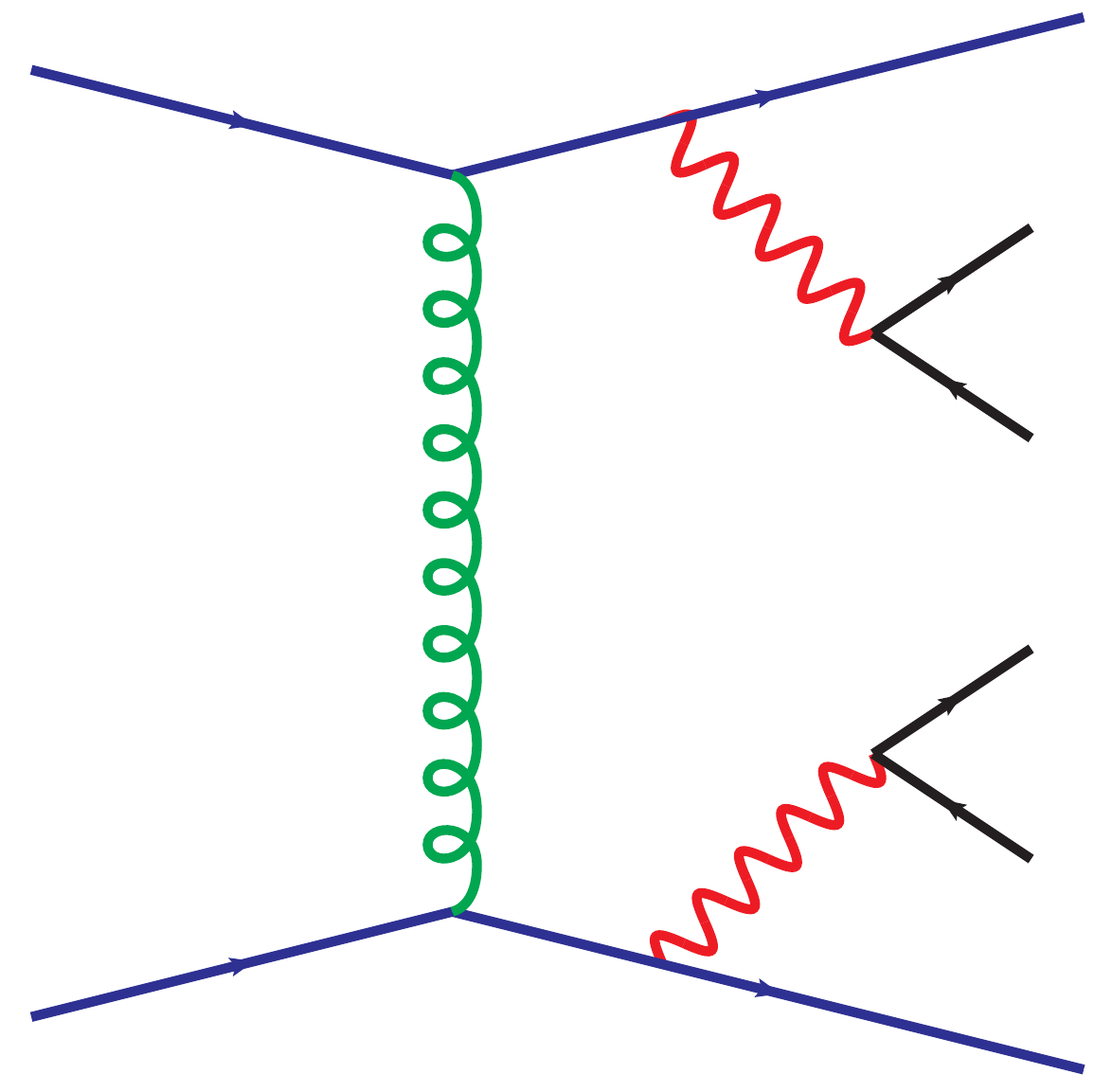} &
\includegraphics[width=0.27\textwidth]{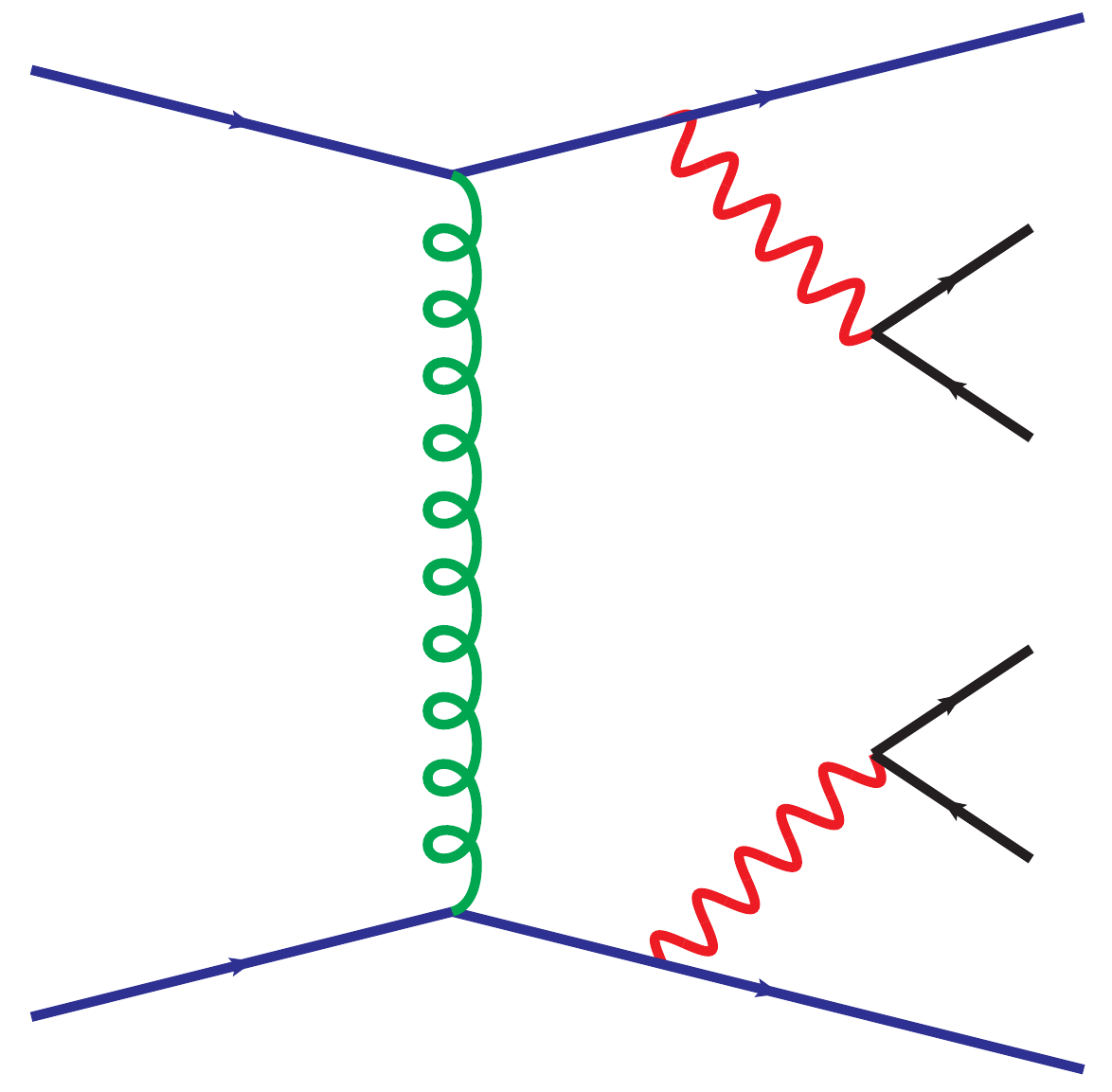} \\
(d) & (e) & (f) \\
\end{tabular}
\end{center}
\caption{Categories of Feynman diagrams that enter a full calculation of.vector boson scattering:
(a),(b)~vector boson scattering or emission through weak interactions;
(c),(d)~Higgs boson production or exchange; 
(e),(f)~vector boson production through strong interactions.}
\label{Fig:VBSdiags}
\end{figure}

In the first case, a class of processes that has attracted theoretical interest for a long
time~\cite{Lee:1977eg} is vector boson scattering.   At a hadron collider such as the LHC,
these processes can be pictured as the emission of a vector boson from each incoming parton,
with the bosons subsequently interacting through the weak coupling and two bosons emerging. 
This picture corresponds to the Feynman diagram shown in  Figure~\ref{Fig:VBSdiags} (a) ,
with the final state consisting of a pair of bosons and two (mostly forward) jets.  However
it should be noted that a full gauge-invariant treatment of this process also includes diagrams
of the form shown in Figure~\ref{Fig:VBSdiags} (b).  These enter at the same order (${\cal O}(\alpha^3)$)
but contain no gauge boson self-interactions.  In addition there are also contributions that involve the Higgs
boson (Figure~\ref{Fig:VBSdiags}, (c) and (d)).  In the Higgs boson resonance region the contribution of diagrams
of type (c) might normally be accounted for separately, to a very good approximation, as the vector boson fusion process.
Finally, there are also QCD-induced contributions
(${\cal O}(\alpha_s \alpha^2)$) such as those shown as (e) and (f) of Figure~\ref{Fig:VBSdiags}. At the very least
these diagrams constitute a non-negligible background.  In the case of contributions such as (e), which share the same
external particles as those of (a)--(d), the QCD and electroweak contributions interfere and a strict separation between
the two types of contribution is not meaningful.

It is interesting to note that contribution (d) means that it is possible to retain sensitivity to Higgs boson
couplings even without the presence of
an $s$-channel resonance, for instance in the case of like-sign $W$-boson production.  In fact this
can be exploited to study off-shell Higgs boson couplings in a similar manner as for inclusive diboson
production discussed above.  Although the rate is much smaller in vector boson scattering channels,
they are tree-level processes and therefore theoretically simpler.  These channels also have a greater
sensitivity at high energies.  The like-sign $W$ channel is particularly promising since the signal
suffers from much smaller QCD-induced backgrounds.  For instance, Ref.~\cite{Campbell:2015vwa} uses
ATLAS evidence for $W^\pm W^\pm$ production~\cite{Aad:2014zda} to obtain a weak bound on the width
of the Higgs boson.

\begin{figure}
\begin{center}
\includegraphics[width=0.49\textwidth]{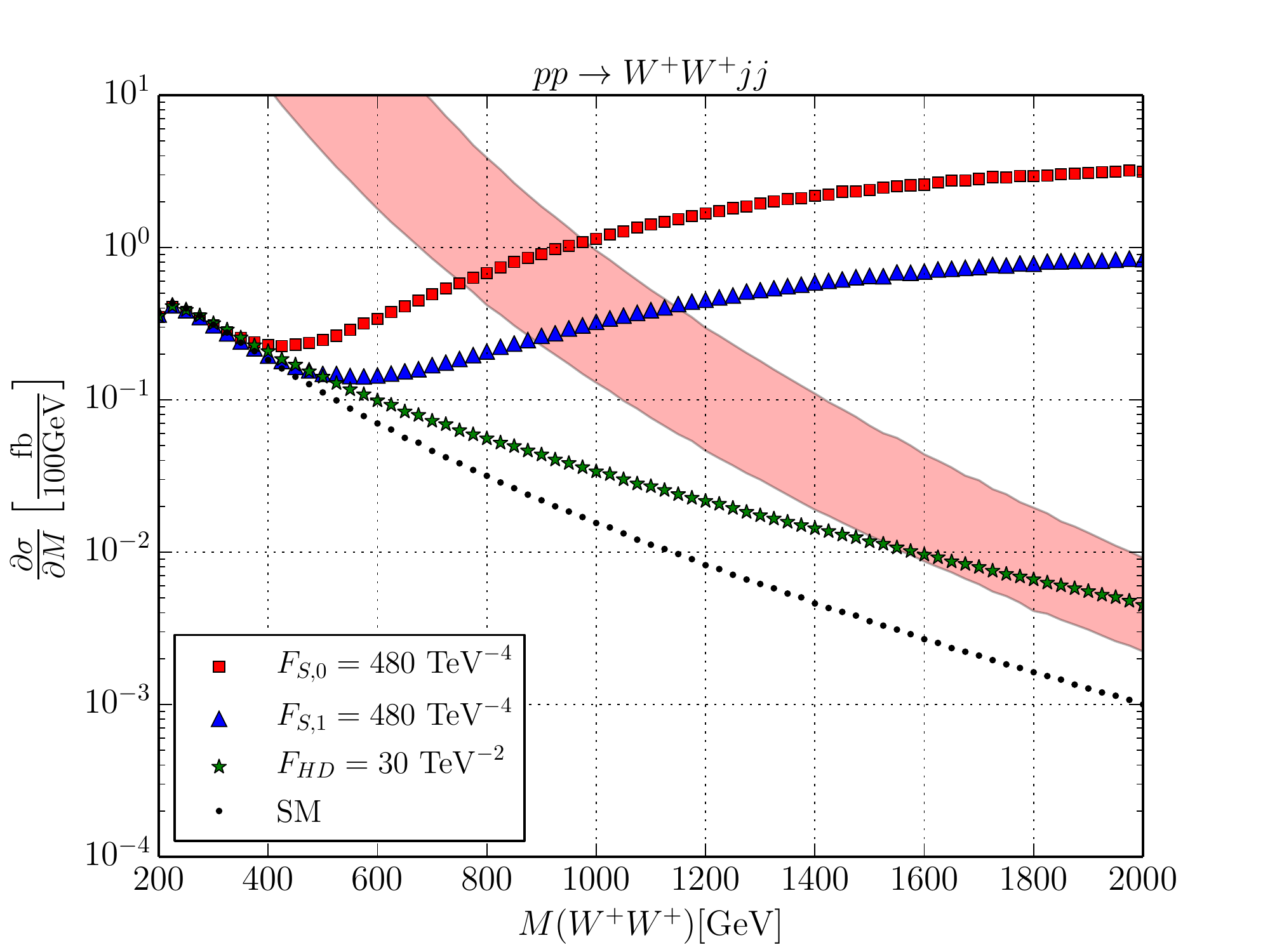}
\includegraphics[width=0.49\textwidth]{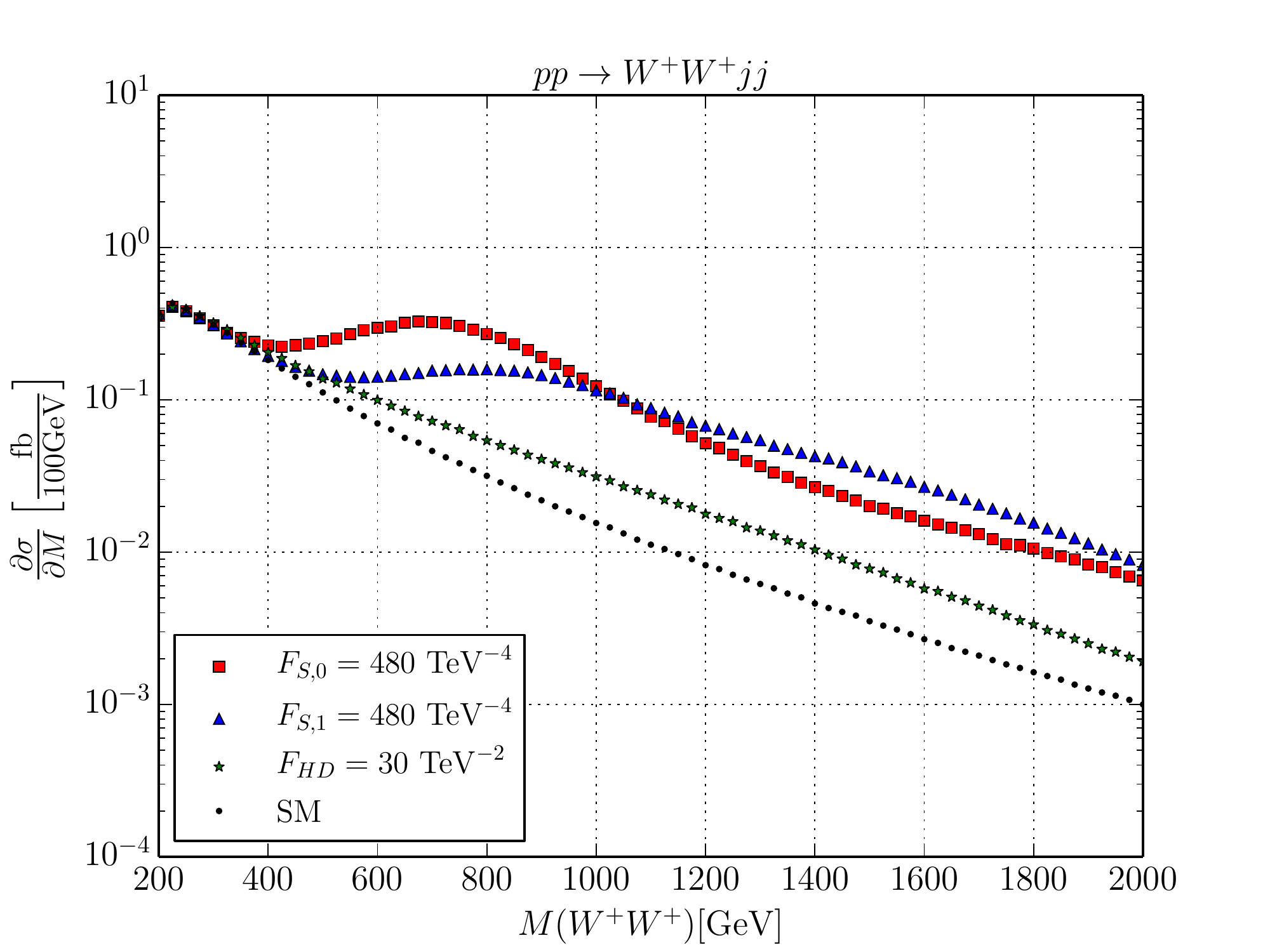}
\end{center}
\caption{Cross-sections for $W^+W^-jj$ production with the addition of the additional interactions
shown in Eq.~(\protect\ref{Eq:EFT}). Unitarity is violated for the values of the coupling strengths
that cross the pink band (left).  After unitarization the high-energy growth is tamed (right).}
\label{Fig:EFT}
\end{figure}

In the last few years there has been much work on developing tools that provide
concrete predictions for signals of vector boson scattering in extensions of the SM.
For instance, the VBFNLO program provides a full suite of predictions for these processes
at NLO, including vector boson decays, off-shell effects and anomalous
couplings~\cite{Campanario:2015vqa}.\footnote{The VBFNLO program was also discussed in the contribution of R. Roth,
{\em ``Anomalous couplings in WZ production beyond NLO QCD''}.}
Other recent work has focussed on extensions that include an additional Higgs singlet or new
resonances~\cite{Ballestrero:2015jca,Kilian:2015opv}.  An alternative approach is to introduce
an effective field theory (EFT) that contains higher-dimension operators that
modify the interactions that occur in vector boson scattering processes.\footnote{For further
details see the contribution of M. Sekulla, {\em ``Effective Field Theory and Unitarity in Vector Boson Scattering''}.}
For example, the study in Ref.~\cite{Kilian:2014zja} has performed a detailed analysis of the effect of additional
terms in the Lagrangian of the form,
\newcommand{\tr}[1]{{\mbox {tr}}\left[#1\right]}
\newcommand{\LL}{\mathcal{L}}
\newcommand{\vH}{\mathbf{H}}
\newcommand{\vD}{\mathbf{D}}
\begin{eqnarray}
\LL_{HD} &=& F_{HD}\;
  \tr{{\vH^\dagger\vH}- \frac{v^2}{4}}\cdot
  \tr{\left (\vD_\mu \vH \right )^\dagger \left (\vD^\mu \vH \right )} \nonumber \\
\LL_{S,0}&=& F_{S,0}\;
  \tr{ \left ( \vD_\mu \vH \right )^\dagger \vD_\nu \vH}
	\cdot \tr{ \left ( \vD^\mu \vH \right )^\dagger \vD^\nu \vH}  \label{Eq:EFT} \\
\LL_{S,1}&=& F_{S,1}\;
  \tr{ \left ( \vD_\mu \vH \right )^\dagger \vD^\mu \vH}
	\cdot \tr{ \left ( \vD_\nu \vH \right )^\dagger \vD^\nu \vH} \nonumber
\end{eqnarray}
These interactions lead to bad high-energy behavior, as shown in Figure~\ref{Fig:EFT} (left) for the
example of $W^+W^+jj$ production at the 14 TeV LHC.  The dimension-six operator
$\LL_{HD}$, which modifies $HW^+W^-$ and $HZZ$ interactions, only violates unitarity at very high
energies, while the other dimension-eight operators (that give rise to modified quartic couplings)
do not give reliable predictions at all.
One solution is provided by a $T$-matrix unitarization procedure~\cite{Kilian:2014zja} that tames the
growth of the prediction at high energy, as indicated in Figure~\ref{Fig:EFT} (right).  In the unitarized theory
there are still pronounced deviations from the SM expectation, but the rates fall with
energy in the same way as in the SM.  Understanding the validity of the EFT approach represented
by the operators in Eq.~(\ref{Eq:EFT}), together with the limitations and uncertainties associated
with unitarization procedures such as the one advocated in Ref.~\cite{Kilian:2014zja}, will be
crucial when interpreting future LHC data on vector boson scattering.

\begin{figure}
\begin{center}
\includegraphics[width=0.5\textwidth]{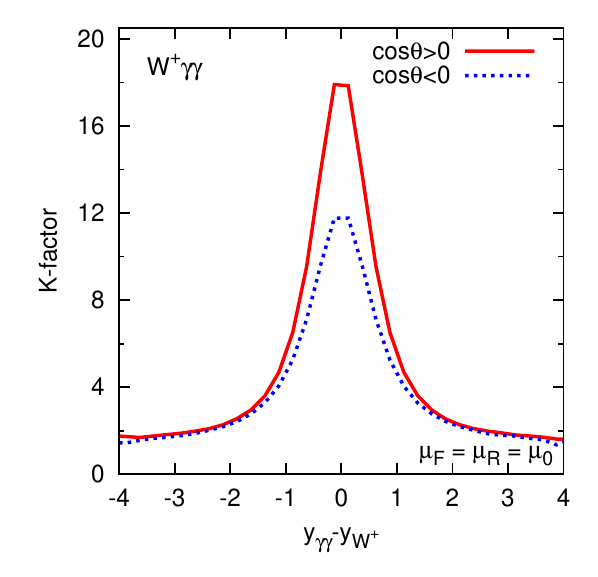}
\end{center}
\caption{The NLO $K$-factor for $W^+\gamma\gamma$ production, as a function of the rapidity
difference between the $\gamma\gamma$ system and the $W^+$-boson, for photons in the
same ($\cos\theta > 0$, solid) or opposite ($\cos\theta < 0$, dashed) hemispheres.  Figure
taken from Ref.~\cite{Bozzi:2011wwa}.}
\label{Fig:Wgamgam}
\end{figure}

The production of three or more bosons represents an alternative means of probing anomalous quartic
gauge boson couplings.  This is an avenue that is only just beginning to be explored at the LHC since
the rate for such processes is at the level of a few tens of femtobarns, after branching ratios into
experimentally-feasible final states and cuts are applied.  For example, the $W\gamma\gamma$ process
is one of the most accessible channels and, despite strong evidence,  it has so far not been
conclusively observed (with more than $5\sigma$ significance) by either ATLAS or
CMS~\cite{Aad:2015uqa,CMS:2016dzs}.  From a theoretical point of view this process is interesting
because it features the well-known effect of a radiation zero~\cite{Mikaelian:1979nr}.  This effect results in
a dip in the distribution of $y(\gamma\gamma) - y(W)$, where $y(\gamma\gamma)$ and $y(W)$ are the
rapidities of the $\gamma\gamma$ system and $W$-boson respectively.  At NLO this dip is filled-in,
particularly by $qg$ partonic channels that enter for the first time, leading to huge $K$-factors
in this region.  This is illustrated in Figure~\ref{Fig:Wgamgam} and results in an enhancement of the
cross-section for this process by a factor of over three at NLO.

Finally, an important background for studies of top quark production, as well as searches for
new physics, is vector boson production in association with jets.  The last year has seen
calculations of $W^+W^-+$jet~\cite{Li:2015ura} and $ZZ+$jet~\cite{Yong:2016njr} production that include
both QCD and EW effects to NLO.  These effects are particularly important to account for at high
invariant masses, where searches for new physics or anomalous couplings are typically focussed.
In the case of multiple-jet production, the current frontier of perturbative calculations has been
pushed as far as $W^+W^- + 3$~jet final states using the combined power of the BlackHat and
SHERPA codes~\cite{Cordero:2015hem}.

\section{Summary}

For electroweak processes precision is paramount, both for SM measurements and for probes of
subtle effects of new physics.  For final states with percent-level experimental precision,
such as vector boson and $Z+$jet production, theoretical predictions are under
control with similar accuracy.  This has been achieved through higher-order calculations
that are accurate to NNLO in QCD and NLO in electroweak effects, and by the development of
new approaches to combining such
effects with a parton shower.   With more jets in the final state both experiment and
theory are a little less pinned-down.  Nevertheless, with more data to come and many of
the theoretical approaches beginning to be better-understood, prospects for precision
probes of the electroweak sector have never been brighter.

\acknowledgments
I thank the organizers for the opportunity to speak at this conference, which
was splendidly coordinated and full of excellent and stimulating talks.
Fermilab is operated by Fermi Research Alliance, LLC under Contract No. DE-AC02-07CH11359
with the United States Department of Energy.

\bibliographystyle{JHEP}
\bibliography{proc}

\providecommand{\href}[2]{#2}\begingroup\raggedright\begin{thebibliography}{10}

\bibitem{Lee:1977eg}
B.~W. Lee, C.~Quigg and H.~B. Thacker, \emph{{Weak Interactions at Very
  High-Energies: The Role of the Higgs Boson Mass}},
  \href{http://dx.doi.org/10.1103/PhysRevD.16.1519}{\emph{Phys. Rev.} {\bf D16}
  (1977) 1519}.

\bibitem{CMSxsecs}
``Summary of {CMS} cross section measurements.''
  {https://twiki.cern.ch/twiki/bin/view/CMSPublic/PhysicsResultsCombined}.

\bibitem{Hoeche:2014aia}
S.~Hoeche, Y.~Li and S.~Prestel, \emph{{Drell-Yan lepton pair production at
  NNLO QCD with parton showers}},
  \href{http://dx.doi.org/10.1103/PhysRevD.91.074015}{\emph{Phys. Rev.} {\bf
  D91} (2015) 074015}, [\href{http://arxiv.org/abs/1405.3607}{{\tt
  1405.3607}}].

\bibitem{Karlberg:2014qua}
A.~Karlberg, E.~Re and G.~Zanderighi, \emph{{NNLOPS accurate Drell-Yan
  production}}, \href{http://dx.doi.org/10.1007/JHEP09(2014)134}{\emph{JHEP}
  {\bf 09} (2014) 134}, [\href{http://arxiv.org/abs/1407.2940}{{\tt
  1407.2940}}].

\bibitem{Alioli:2015toa}
S.~Alioli, C.~W. Bauer, C.~Berggren, F.~J. Tackmann and J.~R. Walsh,
  \emph{{Drell-Yan production at NNLL'+NNLO matched to parton showers}},
  \href{http://dx.doi.org/10.1103/PhysRevD.92.094020}{\emph{Phys. Rev.} {\bf
  D92} (2015) 094020}, [\href{http://arxiv.org/abs/1508.01475}{{\tt
  1508.01475}}].

\bibitem{Boughezal:2016wmq}
R.~Boughezal, J.~M. Campbell, R.~K. Ellis, C.~Focke, W.~Giele, X.~Liu et~al.,
  \emph{{Color singlet production at NNLO in MCFM}},
  \href{http://arxiv.org/abs/1605.08011}{{\tt 1605.08011}}.

\bibitem{Campbell:2016dks}
J.~M. Campbell, D.~Wackeroth and J.~Zhou, \emph{{A Study of Weak Corrections to
  Drell-Yan, Top-quark pair and Di-jet Production at High Energies with MCFM}},
   \href{http://arxiv.org/abs/1608.03356}{{\tt 1608.03356}}.

\bibitem{Li:2012wna}
Y.~Li and F.~Petriello, \emph{{Combining QCD and electroweak corrections to
  dilepton production in FEWZ}},
  \href{http://dx.doi.org/10.1103/PhysRevD.86.094034}{\emph{Phys. Rev.} {\bf
  D86} (2012) 094034}, [\href{http://arxiv.org/abs/1208.5967}{{\tt
  1208.5967}}].

\bibitem{Ciafaloni:1998xg}
P.~Ciafaloni and D.~Comelli, \emph{{Sudakov enhancement of electroweak
  corrections}},
  \href{http://dx.doi.org/10.1016/S0370-2693(98)01541-X}{\emph{Phys. Lett.}
  {\bf B446} (1999) 278--284}, [\href{http://arxiv.org/abs/hep-ph/9809321}{{\tt
  hep-ph/9809321}}].

\bibitem{Bonciani:2016ypc}
R.~Bonciani, S.~Di~Vita, P.~Mastrolia and U.~Schubert, \emph{{Two-Loop Master
  Integrals for the mixed EW-QCD virtual corrections to Drell-Yan scattering}},
   \href{http://arxiv.org/abs/1604.08581}{{\tt 1604.08581}}.

\bibitem{Dittmaier:2015rxo}
S.~Dittmaier, A.~Huss and C.~Schwinn, \emph{{Dominant mixed QCD-electroweak
  $O(\alpha\alpha_s)$ corrections to Drell-Yan processes in the resonance
  region}},
  \href{http://dx.doi.org/10.1016/j.nuclphysb.2016.01.006}{\emph{Nucl. Phys.}
  {\bf B904} (2016) 216--252}, [\href{http://arxiv.org/abs/1511.08016}{{\tt
  1511.08016}}].

\bibitem{Kallweit:2015dum}
S.~Kallweit, J.~M. Lindert, P.~Maierhofer, S.~Pozzorini and M.~Schoenherr,
  \emph{{NLO QCD+EW predictions for V + jets including off-shell vector-boson
  decays and multijet merging}},
  \href{http://dx.doi.org/10.1007/JHEP04(2016)021}{\emph{JHEP} {\bf 04} (2016)
  021}, [\href{http://arxiv.org/abs/1511.08692}{{\tt 1511.08692}}].

\bibitem{Ridder:2015dxa}
A.~Gehrmann-De~Ridder, T.~Gehrmann, E.~W.~N. Glover, A.~Huss and T.~A. Morgan,
  \emph{{Precise QCD predictions for the production of a Z boson in association
  with a hadronic jet}},
  \href{http://dx.doi.org/10.1103/PhysRevLett.117.022001}{\emph{Phys. Rev.
  Lett.} {\bf 117} (2016) 022001}, [\href{http://arxiv.org/abs/1507.02850}{{\tt
  1507.02850}}].

\bibitem{Ridder:2016nkl}
A.~Gehrmann-De~Ridder, T.~Gehrmann, E.~W.~N. Glover, A.~Huss and T.~A. Morgan,
  \emph{{The NNLO QCD corrections to Z boson production at large transverse
  momentum}},  \href{http://arxiv.org/abs/1605.04295}{{\tt 1605.04295}}.

\bibitem{Boughezal:2015dva}
R.~Boughezal, C.~Focke, X.~Liu and F.~Petriello, \emph{{$W$-boson production in
  association with a jet at next-to-next-to-leading order in perturbative
  QCD}}, \href{http://dx.doi.org/10.1103/PhysRevLett.115.062002}{\emph{Phys.
  Rev. Lett.} {\bf 115} (2015) 062002},
  [\href{http://arxiv.org/abs/1504.02131}{{\tt 1504.02131}}].

\bibitem{Boughezal:2015ded}
R.~Boughezal, J.~M. Campbell, R.~K. Ellis, C.~Focke, W.~T. Giele, X.~Liu
  et~al., \emph{{Z-boson production in association with a jet at
  next-to-next-to-leading order in perturbative QCD}},
  \href{http://dx.doi.org/10.1103/PhysRevLett.116.152001}{\emph{Phys. Rev.
  Lett.} {\bf 116} (2016) 152001}, [\href{http://arxiv.org/abs/1512.01291}{{\tt
  1512.01291}}].

\bibitem{Boughezal:2016yfp}
R.~Boughezal, X.~Liu and F.~Petriello, \emph{{A comparison of NNLO QCD
  predictions with 7 TeV ATLAS and CMS data for $V$+jet processes}},
  \href{http://dx.doi.org/10.1016/j.physletb.2016.06.032}{\emph{Phys. Lett.}
  {\bf B760} (2016) 6--13}, [\href{http://arxiv.org/abs/1602.05612}{{\tt
  1602.05612}}].

\bibitem{Catani:2011qz}
S.~Catani, L.~Cieri, D.~de~Florian, G.~Ferrera and M.~Grazzini, \emph{{Diphoton
  production at hadron colliders: a fully-differential QCD calculation at
  NNLO}}, \href{http://dx.doi.org/10.1103/PhysRevLett.108.072001}{\emph{Phys.
  Rev. Lett.} {\bf 108} (2012) 072001},
  [\href{http://arxiv.org/abs/1110.2375}{{\tt 1110.2375}}].

\bibitem{Campbell:2016yrh}
J.~M. Campbell, R.~K. Ellis, Y.~Li and C.~Williams, \emph{{Predictions for
  diphoton production at the LHC through NNLO in QCD}},
  \href{http://arxiv.org/abs/1603.02663}{{\tt 1603.02663}}.

\bibitem{Grazzini:2015nwa}
M.~Grazzini, S.~Kallweit and D.~Rathlev, \emph{{$W\gamma$ and $Z\gamma$
  production at the LHC in NNLO QCD}},
  \href{http://dx.doi.org/10.1007/JHEP07(2015)085}{\emph{JHEP} {\bf 07} (2015)
  085}, [\href{http://arxiv.org/abs/1504.01330}{{\tt 1504.01330}}].

\bibitem{Denner:2014bna}
A.~Denner, S.~Dittmaier, M.~Hecht and C.~Pasold, \emph{{NLO QCD and electroweak
  corrections to $W+\gamma$ production with leptonic W-boson decays}},
  \href{http://dx.doi.org/10.1007/JHEP04(2015)018}{\emph{JHEP} {\bf 04} (2015)
  018}, [\href{http://arxiv.org/abs/1412.7421}{{\tt 1412.7421}}].

\bibitem{Denner:2015fca}
A.~Denner, S.~Dittmaier, M.~Hecht and C.~Pasold, \emph{{NLO QCD and electroweak
  corrections to $Z+\gamma$ production with leptonic Z-boson decays}},
  \href{http://dx.doi.org/10.1007/JHEP02(2016)057}{\emph{JHEP} {\bf 02} (2016)
  057}, [\href{http://arxiv.org/abs/1510.08742}{{\tt 1510.08742}}].

\bibitem{Gehrmann:2014fva}
T.~Gehrmann, M.~Grazzini, S.~Kallweit, P.~Maierhoefer, A.~von Manteuffel,
  S.~Pozzorini et~al., \emph{{$W^+W^-$ Production at Hadron Colliders in Next
  to Next to Leading Order QCD}},
  \href{http://dx.doi.org/10.1103/PhysRevLett.113.212001}{\emph{Phys. Rev.
  Lett.} {\bf 113} (2014) 212001}, [\href{http://arxiv.org/abs/1408.5243}{{\tt
  1408.5243}}].

\bibitem{Grazzini:2016ctr}
M.~Grazzini, S.~Kallweit, S.~Pozzorini, D.~Rathlev and M.~Wiesemann,
  \emph{{$W^+W^-$ production at the LHC: fiducial cross sections and
  distributions in NNLO QCD}},  \href{http://arxiv.org/abs/1605.02716}{{\tt
  1605.02716}}.

\bibitem{Caola:2015rqy}
F.~Caola, K.~Melnikov, R.~Roentsch and L.~Tancredi, \emph{{QCD corrections to
  $W^+W^-$ production through gluon fusion}},
  \href{http://dx.doi.org/10.1016/j.physletb.2016.01.046}{\emph{Phys. Lett.}
  {\bf B754} (2016) 275--280}, [\href{http://arxiv.org/abs/1511.08617}{{\tt
  1511.08617}}].

\bibitem{Biedermann:2016guo}
B.~Biedermann, M.~Billoni, A.~Denner, S.~Dittmaier, L.~Hofer, B.~Jaeger et~al.,
  \emph{{Next-to-leading-order electroweak corrections to $pp \to W^+W^-\to$ 4
  leptons at the LHC}},
  \href{http://dx.doi.org/10.1007/JHEP06(2016)065}{\emph{JHEP} {\bf 06} (2016)
  065}, [\href{http://arxiv.org/abs/1605.03419}{{\tt 1605.03419}}].

\bibitem{Grazzini:2016swo}
M.~Grazzini, S.~Kallweit, D.~Rathlev and M.~Wiesemann, \emph{{$W^{\pm}Z$
  production at hadron colliders in NNLO QCD}},
  \href{http://arxiv.org/abs/1604.08576}{{\tt 1604.08576}}.

\bibitem{Grazzini:2015hta}
M.~Grazzini, S.~Kallweit and D.~Rathlev, \emph{{ZZ production at the LHC:
  fiducial cross sections and distributions in NNLO QCD}},
  \href{http://dx.doi.org/10.1016/j.physletb.2015.09.055}{\emph{Phys. Lett.}
  {\bf B750} (2015) 407--410}, [\href{http://arxiv.org/abs/1507.06257}{{\tt
  1507.06257}}].

\bibitem{Caola:2015psa}
F.~Caola, K.~Melnikov, R.~Roentsch and L.~Tancredi, \emph{{QCD corrections to
  $ZZ$ production in gluon fusion at the LHC}},
  \href{http://dx.doi.org/10.1103/PhysRevD.92.094028}{\emph{Phys. Rev.} {\bf
  D92} (2015) 094028}, [\href{http://arxiv.org/abs/1509.06734}{{\tt
  1509.06734}}].

\bibitem{Biedermann:2016yvs}
B.~Biedermann, A.~Denner, S.~Dittmaier, L.~Hofer and B.~Jaeger,
  \emph{{Electroweak corrections to $pp \to \mu^+\mu^-e^+e^- + X$ at the LHC: a
  Higgs background study}},
  \href{http://dx.doi.org/10.1103/PhysRevLett.116.161803}{\emph{Phys. Rev.
  Lett.} {\bf 116} (2016) 161803}, [\href{http://arxiv.org/abs/1601.07787}{{\tt
  1601.07787}}].

\bibitem{Aad:2016sau}
{\scshape ATLAS} collaboration, G.~Aad et~al., \emph{{Measurements of $Z\gamma$
  and $Z\gamma\gamma$ production in $pp$ collisions at $\sqrt{s}=$ 8 TeV with
  the ATLAS detector}},
  \href{http://dx.doi.org/10.1103/PhysRevD.93.112002}{\emph{Phys. Rev.} {\bf
  D93} (2016) 112002}, [\href{http://arxiv.org/abs/1604.05232}{{\tt
  1604.05232}}].

\bibitem{Bern:2001df}
Z.~Bern, A.~De~Freitas and L.~J. Dixon, \emph{{Two loop amplitudes for gluon
  fusion into two photons}},
  \href{http://dx.doi.org/10.1088/1126-6708/2001/09/037}{\emph{JHEP} {\bf 09}
  (2001) 037}, [\href{http://arxiv.org/abs/hep-ph/0109078}{{\tt
  hep-ph/0109078}}].

\bibitem{Bern:2002jx}
Z.~Bern, L.~J. Dixon and C.~Schmidt, \emph{{Isolating a light Higgs boson from
  the diphoton background at the CERN LHC}},
  \href{http://dx.doi.org/10.1103/PhysRevD.66.074018}{\emph{Phys. Rev.} {\bf
  D66} (2002) 074018}, [\href{http://arxiv.org/abs/hep-ph/0206194}{{\tt
  hep-ph/0206194}}].

\bibitem{Chatrchyan:2014fsa}
{\scshape CMS} collaboration, S.~Chatrchyan et~al., \emph{{Measurement of
  differential cross sections for the production of a pair of isolated photons
  in pp collisions at $\sqrt{s}=7$ TeV}},
  \href{http://dx.doi.org/10.1140/epjc/s10052-014-3129-3}{\emph{Eur. Phys. J.}
  {\bf C74} (2014) 3129}, [\href{http://arxiv.org/abs/1405.7225}{{\tt
  1405.7225}}].

\bibitem{ATLAS-CONF-2015-081}
{ATLAS collaboration}, \emph{{Search for resonances decaying to photon pairs in
  3.2 fb$^{-1}$ of $pp$ collisions at $\sqrt{s}$ = 13 TeV with the ATLAS
  detector}},  Tech. Rep. ATLAS-CONF-2015-081, CERN, Geneva, Dec, 2015.

\bibitem{CMS-PAS-EXO-15-004}
{CMS collaboration}, \emph{{Search for new physics in high mass diphoton events
  in proton-proton collisions at $\sqrt{s} = 13$ TeV}},  Tech. Rep.
  CMS-PAS-EXO-15-004, CERN, Geneva, 2015.

\bibitem{Aad:2016wpd}
{\scshape ATLAS} collaboration, G.~Aad et~al., \emph{{Measurement of total and
  differential $W^+W^-$ production cross sections in proton-proton collisions
  at $\sqrt{s}=$ 8 TeV with the ATLAS detector and limits on anomalous
  triple-gauge-boson couplings}},  \href{http://arxiv.org/abs/1603.01702}{{\tt
  1603.01702}}.

\bibitem{ATLAS:2013gma}
{ATLAS collaboration}, \emph{{Measurement of the total ZZ production cross
  section in proton-proton collisions at $\sqrt{s} = 8$ TeV in 20 fb${}^{-1}$
  with the ATLAS detector}}, .

\bibitem{CMS:2014xja}
{\scshape CMS} collaboration, V.~Khachatryan et~al., \emph{{Measurement of the
  $pp \to ZZ$ production cross section and constraints on anomalous triple
  gauge couplings in four-lepton final states at $\sqrt s=$8 TeV}},
  \href{http://dx.doi.org/10.1016/j.physletb.2016.04.010,
  10.1016/j.physletb.2014.11.059}{\emph{Phys. Lett.} {\bf B740} (2015)
  250--272}, [\href{http://arxiv.org/abs/1406.0113}{{\tt 1406.0113}}].

\bibitem{Kauer:2012hd}
N.~Kauer and G.~Passarino, \emph{{Inadequacy of zero-width approximation for a
  light Higgs boson signal}},
  \href{http://dx.doi.org/10.1007/JHEP08(2012)116}{\emph{JHEP} {\bf 08} (2012)
  116}, [\href{http://arxiv.org/abs/1206.4803}{{\tt 1206.4803}}].

\bibitem{Caola:2013yja}
F.~Caola and K.~Melnikov, \emph{{Constraining the Higgs boson width with ZZ
  production at the LHC}},
  \href{http://dx.doi.org/10.1103/PhysRevD.88.054024}{\emph{Phys. Rev.} {\bf
  D88} (2013) 054024}, [\href{http://arxiv.org/abs/1307.4935}{{\tt
  1307.4935}}].

\bibitem{Glover:1988rg}
E.~N. Glover and J.~van~der Bij, \emph{{$Z$-boson pair production via gluon
  fusion}},
  \href{http://dx.doi.org/10.1016/0550-3213(89)90262-9}{\emph{Nucl.Phys.} {\bf
  B321} (1989) 561}.

\bibitem{Campbell:2013una}
J.~M. Campbell, R.~K. Ellis and C.~Williams, \emph{{Bounding the Higgs width at
  the LHC using full analytic results for $gg -> e^- e^+ \mu^- \mu^+$}},
  \href{http://dx.doi.org/10.1007/JHEP04(2014)060}{\emph{JHEP} {\bf 04} (2014)
  060}, [\href{http://arxiv.org/abs/1311.3589}{{\tt 1311.3589}}].

\bibitem{Aad:2015xua}
{\scshape ATLAS} collaboration, G.~Aad et~al., \emph{{Constraints on the
  off-shell Higgs boson signal strength in the high-mass $ZZ$ and $WW$ final
  states with the ATLAS detector}},
  \href{http://dx.doi.org/10.1140/epjc/s10052-015-3542-2}{\emph{Eur. Phys. J.}
  {\bf C75} (2015) 335}, [\href{http://arxiv.org/abs/1503.01060}{{\tt
  1503.01060}}].

\bibitem{Li:2015jva}
C.~S. Li, H.~T. Li, D.~Y. Shao and J.~Wang, \emph{{Soft gluon resummation in
  the signal-background interference process of $gg(\to h^{∗}) \to ZZ$}},
  \href{http://dx.doi.org/10.1007/JHEP08(2015)065}{\emph{JHEP} {\bf 08} (2015)
  065}, [\href{http://arxiv.org/abs/1504.02388}{{\tt 1504.02388}}].

\bibitem{Melnikov:2015laa}
K.~Melnikov and M.~Dowling, \emph{{Production of two Z-bosons in gluon fusion
  in the heavy top quark approximation}},
  \href{http://dx.doi.org/10.1016/j.physletb.2015.03.030}{\emph{Phys. Lett.}
  {\bf B744} (2015) 43--47}, [\href{http://arxiv.org/abs/1503.01274}{{\tt
  1503.01274}}].

\bibitem{Caola:2016trd}
F.~Caola, M.~Dowling, K.~Melnikov, R.~R{\"o}ntsch and L.~Tancredi, \emph{{QCD
  corrections to vector boson pair production in gluon fusion including
  interference effects with off-shell Higgs at the LHC}},
  \href{http://dx.doi.org/10.1007/JHEP07(2016)087}{\emph{JHEP} {\bf 07} (2016)
  087}, [\href{http://arxiv.org/abs/1605.04610}{{\tt 1605.04610}}].

\bibitem{Campbell:2016ivq}
J.~M. Campbell, R.~K. Ellis, M.~Czakon and S.~Kirchner, \emph{{Two loop
  correction to interference in $gg \to ZZ$}},
  \href{http://dx.doi.org/10.1007/JHEP08(2016)011}{\emph{JHEP} {\bf 08} (2016)
  011}, [\href{http://arxiv.org/abs/1605.01380}{{\tt 1605.01380}}].

\bibitem{Campbell:2015vwa}
J.~M. Campbell and R.~K. Ellis, \emph{{Higgs Constraints from Vector Boson
  Fusion and Scattering}},
  \href{http://dx.doi.org/10.1007/JHEP04(2015)030}{\emph{JHEP} {\bf 04} (2015)
  030}, [\href{http://arxiv.org/abs/1502.02990}{{\tt 1502.02990}}].

\bibitem{Aad:2014zda}
{\scshape ATLAS} collaboration, G.~Aad et~al., \emph{{Evidence for Electroweak
  Production of $W^{\pm}W^{\pm}jj$ in $pp$ Collisions at $\sqrt{s}=8$ TeV with
  the ATLAS Detector}},
  \href{http://dx.doi.org/10.1103/PhysRevLett.113.141803}{\emph{Phys. Rev.
  Lett.} {\bf 113} (2014) 141803}, [\href{http://arxiv.org/abs/1405.6241}{{\tt
  1405.6241}}].

\bibitem{Campanario:2015vqa}
F.~Campanario, M.~Kerner, L.~D. Ninh, M.~Rauch, R.~Roth and D.~Zeppenfeld,
  \emph{{NLO corrections to processes with electroweak bosons at hadron
  colliders}},
  \href{http://dx.doi.org/10.1016/j.nuclphysbps.2015.03.019}{\emph{Nucl. Part.
  Phys. Proc.} {\bf 261-262} (2015) 268--307}.

\bibitem{Ballestrero:2015jca}
A.~Ballestrero and E.~Maina, \emph{{Interference Effects in Higgs production
  through Vector Boson Fusion in the Standard Model and its Singlet
  Extension}}, \href{http://dx.doi.org/10.1007/JHEP01(2016)045}{\emph{JHEP}
  {\bf 01} (2016) 045}, [\href{http://arxiv.org/abs/1506.02257}{{\tt
  1506.02257}}].

\bibitem{Kilian:2015opv}
W.~Kilian, T.~Ohl, J.~Reuter and M.~Sekulla, \emph{{Resonances at the LHC
  beyond the Higgs boson: The scalar/tensor case}},
  \href{http://dx.doi.org/10.1103/PhysRevD.93.036004}{\emph{Phys. Rev.} {\bf
  D93} (2016) 036004}, [\href{http://arxiv.org/abs/1511.00022}{{\tt
  1511.00022}}].

\bibitem{Kilian:2014zja}
W.~Kilian, T.~Ohl, J.~Reuter and M.~Sekulla, \emph{{High-Energy Vector Boson
  Scattering after the Higgs Discovery}},
  \href{http://dx.doi.org/10.1103/PhysRevD.91.096007}{\emph{Phys. Rev.} {\bf
  D91} (2015) 096007}, [\href{http://arxiv.org/abs/1408.6207}{{\tt
  1408.6207}}].

\bibitem{Aad:2015uqa}
{\scshape ATLAS} collaboration, G.~Aad et~al., \emph{{Evidence of
  $W\gamma\gamma$ Production in pp Collisions at $\sqrt{s}=8$~TeV and Limits on
  Anomalous Quartic Gauge Couplings with the ATLAS Detector}},
  \href{http://dx.doi.org/10.1103/PhysRevLett.115.031802}{\emph{Phys. Rev.
  Lett.} {\bf 115} (2015) 031802}, [\href{http://arxiv.org/abs/1503.03243}{{\tt
  1503.03243}}].

\bibitem{CMS:2016dzs}
{CMS collaboration}, \emph{{Measurements of The
  $pp\to\mathrm{W}^{\pm}\gamma\gamma$ and $pp\to\mathrm{Z}\gamma\gamma$ Cross
  Sections and Limits on Dimension-8 Effective Anomalous Gauge Couplings at
  $\sqrt{s} = 8~\mathrm{TeV}$}},  Tech. Rep. CMS-PAS-SMP-15-008, CERN, 2016.

\bibitem{Mikaelian:1979nr}
K.~Mikaelian, M.~Samuel and D.~Sahdev, \emph{{The magnetic moment of weak
  bosons produced in $p p$ and $p \bar{p}$ collisions}},
  \href{http://dx.doi.org/10.1103/PhysRevLett.43.746}{\emph{Phys.Rev.Lett.}
  {\bf 43} (1979) 746}.

\bibitem{Li:2015ura}
W.-H. Li, R.-Y. Zhang, W.-G. Ma, L.~Guo, X.-Z. Li and Y.~Zhang, \emph{{NLO QCD
  and electroweak corrections to $WW+$jet production with leptonic $W$ -boson
  decays at LHC}},
  \href{http://dx.doi.org/10.1103/PhysRevD.92.033005}{\emph{Phys. Rev.} {\bf
  D92} (2015) 033005}, [\href{http://arxiv.org/abs/1507.07332}{{\tt
  1507.07332}}].

\bibitem{Yong:2016njr}
Y.~Wang, R.-Y. Zhang, W.-G. Ma, X.-Z. Li and L.~Guo, \emph{{QCD and electroweak
  corrections to ZZ+jet production with Z -boson leptonic decays at the LHC}},
  \href{http://dx.doi.org/10.1103/PhysRevD.94.013011}{\emph{Phys. Rev.} {\bf
  D94} (2016) 013011}, [\href{http://arxiv.org/abs/1604.04080}{{\tt
  1604.04080}}].

\bibitem{Cordero:2015hem}
F.~Febres~Cordero, P.~Hofmann and H.~Ita, \emph{{$W^+W^-$ + 3 Jet Production at
  the Large Hadron Collider in NLO QCD}},
  \href{http://arxiv.org/abs/1512.07591}{{\tt 1512.07591}}.

\bibitem{Bozzi:2011wwa}
G.~Bozzi, F.~Campanario, M.~Rauch and D.~Zeppenfeld, \emph{{$W^{+-}\gamma
  \gamma$ production with leptonic decays at NLO QCD}},
  \href{http://dx.doi.org/10.1103/PhysRevD.83.114035}{\emph{Phys. Rev.} {\bf
  D83} (2011) 114035}, [\href{http://arxiv.org/abs/1103.4613}{{\tt
  1103.4613}}].

\end{thebibliography}\endgroup

\end{document}